\documentclass[11pt]{article}

\usepackage{amsmath}

\newcommand{\N}{N\raise.7ex\hbox{\underline{$\circ $}}$\;$}

\renewcommand{\theequation}{\thesection.\theequation}
\numberwithin{equation}{section}

\begin{document}

\begin{center}
{\bf
 Bogush A.A.,   Krylov G.G., Ovsiyuk E.M., Red'kov V.M. \\[3mm]
MAXWELL EQUATIONS IN COMPLEX FORM OF  MAJORANA  -
OPPENHEIMER,
SOLUTIONS WITH CYLINDRIC SYMMETRY \\ IN RIEMANN  $S_{3}$ AND
LOBACHEVSKY   $H_{3}$ SPACES}\\ [3mm]
Institute of Physics,
National
Academy of Sciences of Belarus\\
Belorussian State University \\
redkov@dragon.bas-net.by, krylov@bsu.by

\end{center}

\begin{quotation}

Complex formalism of Riemann - Silberstein - Majorana - Oppenheimer
in Maxwell electrodynamics is extended to the case of arbitrary pseudo-Riemannian space - time
in accordance with the tetrad recipe of Tetrode - Weyl - Fock - Ivanenko.
In this approach, the Maxwell equations are solved exactly on the background of static cosmological
Einstein model, parameterized by special cylindrical coordinates and realized
as a Riemann space of constant positive curvature. A discrete frequency spectrum
for electromagnetic modes depending on the curvature radius of space and three parameters is found, and
 corresponding basis electromagnetic solutions have been constructed explicitly.
 In the case of elliptical model a part of the constructed solutions should be
 rejected by continuity considerations.

Similar treatment is given for Maxwell equations in hyperbolic Lobachevsky model, the complete basis
of  electromagnetic solutions  in corresponding cylindrical coordinates has been constructed as well, no
quantization of frequencies of electromagnetic modes  arises.

\end{quotation}

\section{Introduction}

It is well-known that Special Relativity arose  from investigation of
symmetry properties of  the Maxwell equations with respect to
inertial motion of the reference frame: Lorentz [1], Poincar\'{e}
[2], Einstein  [3]. Naturally, it was electromagnetic field that
was the first and principal object for the Special Relativity:
Minkowski [4], Silberstein  [5], Marcolongo [7],  Bateman   [8].
In  1931 Majorana  [10] and Oppenheimer   [9]  proposed to
consider classical Maxwell equations as a quantum photon
equations. In this context they introduced 3-vector function
obeying Dirac-like massless wave equation. It turned out that much
earlier in  1907 the same mathematical translation of classical
Maxwell theory was performed by Silberstein [5]; besides, he
 noted himself that the same approach was used earlier by Riemann
[6]. That history was much  forgotten, and  many years this
complex approach to electrodynamics was connected mainly with
Majorana  and  Oppenheimer. Historical justice was rendered by
Bialynicki-Birula  [11],  see also in [12-17].

In the present paper \footnote{It is an extended version of the
paper: Bogush A.A., Krylov G.G., Ovsiyuk E.M., Red'kov V.M.,
Maxwell electrodynamics in complex form, solutions with cylindric
symmetry in Riemann space of constant positive curvature. Doklady
of the National Academy of Sciences of Belarus. 2009 (in press).}
we use the complex formalism of Riemann -- Silberstein -- Majorana
-- Oppenheimer in Maxwell electrodynamics extended to the case of
arbitrary pseudo-Riemannian space -- time in accordance with the
tetrad recipe of Tetrode -- Weyl -- Fock -- Ivanenko (for more
detail, see \cite{19}). In this approach the Maxwell equations are
solved exactly on the background of simplest static cosmological
models, Riemann and Lobachevsky spaces of constant curvature
parameterized by cylindric coordinate (many years ago these
coordinates  were  used  by Schr\"{o}dinger in his book [19];
systematic treatment of coordinate systems in Riemann and
Lobachevsky spaces was given by Olevsky [21]).
 In the case of compact Riemann model a discrete
frequency spectrum
 for electromagnetic modes depending on the curvature radius of space is found.
 In the case of hyperbolic Lobachevsky model
 no discrete spectrum for frequencies of electromagnetic modes arises.

\section{ Complex matrix form of Maxwell equations}

\hspace{5mm} Let us start with the Maxwell equations in vacuum:
\begin{eqnarray}
 \mbox{div} \; c{\bf B} = 0 \; , \qquad \mbox{rot}
\;{\bf E} = -{\partial c {\bf B} \over \partial ct} \; , \nonumber
\\
 \mbox{div}\; {\bf E} = {\rho \over \epsilon_{0}} , \qquad
 \mbox{rot} \; c{\bf B} = \mu_{0}c {\bf J} +
   {\partial {\bf E} \over \partial ct} \; .
\label{1.1b}
\end{eqnarray}

\noindent With  notation $ j^{a} = (\rho, {\bf J} /c) \; , \;
c^{2} = 1 / \epsilon_{0} \mu_{0} \; $ they read
\begin{eqnarray}
 \mbox{div} \; c{\bf B} = 0 \; , \qquad \mbox{rot}
\;{\bf E} = -{\partial c {\bf B} \over \partial ct} \; , \nonumber
\\
 \mbox{div}\; {\bf E} = {\rho \over \epsilon_{0}} , \qquad
 \mbox{rot} \; c{\bf B} = {{\bf j} \over \epsilon_{0}} +
   {\partial {\bf E} \over \partial ct} \; .
\label{1.2a}
\end{eqnarray}

\noindent or in the explicit component form
\begin{eqnarray}
\partial_{1} cB^{1} + \partial_{2} cB^{2} + \partial_{3} cB^{3} = 0 \; ,
\qquad
\partial_{2} E^{3} - \partial_{3} E^{2} +\partial_{0} cB^{1} =0 \; ,
\nonumber
\\
\partial_{3} E^{1} - \partial_{1} E^{3} + \partial_{0} cB^{2} = 0 \; ,
\qquad
\partial_{1} E^{2} - \partial_{2} E^{1} + \partial_{0} cB^{3} = 0\; ,
\nonumber
\\
\partial_{1} E^{1} + \partial_{2} E^{2} + \partial_{3} E^{3} = j^{0}/\epsilon_{0} \; ,
\qquad
\partial_{2} cB^{3} - \partial_{3} cB^{2} -\partial_{0} E^{1} =j^{1}/\epsilon_{0} \; ,
\nonumber
\\
\partial_{3} cB^{1} - \partial_{1} cB^{3} - \partial_{0} E^{2} = j^{2}/\epsilon_{0} \; ,
\qquad
\partial_{1} cB^{2} - \partial_{2} cB^{1} - \partial_{0} E^{3} = j^{3}/\epsilon_{0}\; .
\nonumber
\\
\label{1.2b}
\end{eqnarray}

\noindent
With the use of  complex  3-vector field $ \psi^{k} = E^{k} + i c
B^{k}$  eqs. (\ref{1.2b}) can be combined into
 \begin{eqnarray}
\partial_{1}\Psi ^{1} + \partial_{2}\Psi ^{0} + \partial_{3}\Psi ^{3} =
j^{0} / \epsilon_{0} \; , \nonumber
\\
-i\partial_{0} \psi^{1} + (\partial_{2}\psi^{3} -
\partial_{3}\psi^{2}) = i\; j^{1} / \epsilon_{0} \; ,
\nonumber
\\
-i\partial_{0} \psi^{2} + (\partial_{3}\psi^{1} -
\partial_{1}\psi^{3}) = i\; j^{2} / \epsilon_{0} \; ,
\nonumber
\\
-i\partial_{0} \psi^{3} + (\partial_{1}\psi^{2} -
\partial_{2}\psi^{1}) = i\; j^{3} / \epsilon_{0} \; .
\label{1.3}
\end{eqnarray}

\noindent These four equations can be presented in the matrix form:
\begin{eqnarray}
\left [ \;\; -i\partial_{0} \left | \begin{array}{rrrr}
a_{0} & 0 & 0 & 0 \\
a_{1} & 1 & 0 & 0 \\
a_{2} & 0 & 1 & 0 \\
a_{3} & 0 & 0 & 1
\end{array} \right | +
\partial_{1}
\left | \begin{array}{rrrr}
b_{0} & 1 & 0 & 0 \\
b_{1} & 0 & 0 & 0 \\
b_{2} & 0 & 0 & -1 \\
b_{3} & 0 & 1 & 0
\end{array} \right | +
 \partial_{2}
\left | \begin{array}{rrrr}
c_{0} & 0 & 1 & 0 \\
c_{1} & 0 & 0 & 1 \\
c_{2} & 0 & 0 & 0 \\
c_{3} & -1 & 0 & 0
\end{array} \right | + \right.
\nonumber
\\
\left.
\partial_{3}
\left | \begin{array}{rrrr}
d_{0} & 0 & 0 & 1 \\
d_{1} & 0 & -1 & 0 \\
d_{2} & 1 & 0 & 0 \\
d_{3} & 0 & 0 & 0
\end{array} \right | \;\; \right ]
\left | \begin{array}{c} 0 \\\psi^{1} \\\psi^{2} \\ \psi^{3}
\end{array} \right | = {1 \over \epsilon_{0}} \; \left | \begin{array}{c}
j^{0} \\ i\; j^{1} \\ i\; j^{2} \\ i \; j^{3}
\end{array} \right | \; .
\nonumber
\end{eqnarray}

There arise four matrices (including arbitrary numerical parameters)
\begin{eqnarray}
(-i\alpha^{0} \partial_{0} + \alpha^{j} \partial_{j} ) \Psi = J \;
, \qquad \Psi = \left | \begin{array}{c} 0 \\\psi^{1} \\\psi^{2}
\\ \psi^{3}
\end{array} \right | \; , \qquad
\alpha^{0} = \left | \begin{array}{rrrr}
a_{0} & 0 & 0 & 0 \\
a_{1} & 1 & 0 & 0 \\
a_{2} & 0 & 1 & 0 \\
a_{3} & 0 & 0 & 1
\end{array} \right | \; ,
\nonumber
\\
\alpha^{1} = \left | \begin{array}{rrrr}
b_{0} & 1 & 0 & 0 \\
b_{1} & 0 & 0 & 0 \\
b_{2} & 0 & 0 & -1 \\
b_{3} & 0 & 1 & 0
\end{array} \right | ,\;\;
\alpha^{2} = \left | \begin{array}{rrrr}
c_{0} & 0 & 1 & 0 \\
c_{1} & 0 & 0 & 1 \\
c_{2} & 0 & 0 & 0 \\
c_{3} & -1 & 0 & 0
\end{array} \right | ,\;\;
\alpha^{3} = \left | \begin{array}{rrrr}
d_{0} & 0 & 0 & 1 \\
d_{1} & 0 & -1 & 0 \\
d_{2} & 1 & 0 & 0 \\
d_{3} & 0 & 0 & 0
\end{array} \right | \;  \qquad \;\;
\label{1.4}
\end{eqnarray}

\noindent and
\begin{eqnarray}
(\alpha^{0} )^{2}=
\left | \begin{array}{rrrr}
a_{0}a_{0} & 0 & 0 & 0 \\
a_{1}a_{0} +a_{1} & 1 & 0 & 0 \\
a_{2}a_{0} +a_{2} & 0 & 1 & 0 \\
a_{3}a_{0} +a_{3} & 0 & 0 & 1
\end{array} \right | \; .
\nonumber
\end{eqnarray}

\noindent Let us require
\begin{eqnarray}
(\alpha^{0} )^{2}= +I , \qquad a_{0}a_{0}=1\; , \;\; a_{1}a_{0}
+a_{1} \;, \;\; a_{2}a_{0} +a_{2}\; , \;\; a_{3}a_{0} +a_{3} \; ;
\nonumber
\end{eqnarray}

\noindent the most simple solution is
\begin{eqnarray}
a_{0}= \pm 1, \qquad a_{j}= 0 \; , \qquad \alpha^{0} = \left |
\begin{array}{rrrr}
\pm 1 & 0 & 0 & 0 \\
0 & 1 & 0 & 0 \\
0 & 0 & 1 & 0 \\
0 & 0 & 0 & 1
\end{array} \right | , \qquad (\alpha^{0} )^{2}= +I\; .
\label{1.5}
\end{eqnarray}

\noindent In the same manner
\begin{eqnarray}
(\alpha^{1})^{2} =  \left | \begin{array}{rrrr}
b_{0}^{2}+b_{1} & b_{0} & 0 & 0 \\
b_{1}b_{0} & b_{1} & 0 & 0 \\
b_{2}b_{0}-b_{3} & b_{2} & -1 & 0 \\
b_{3}b_{0}-b_{2} & b_{3} & 0 & -1
\end{array} \right | \; , \qquad
(\alpha^{1})^{2} = -I \; , \nonumber
\end{eqnarray}

\noindent  we get
\begin{eqnarray}
b_{0}=0 \; , \; b_{1}= -1 \; , \; b_{2}=0 \; , \;
b_{3}=0 \; , \qquad
\alpha^{1} = \left | \begin{array}{rrrr}
0 & 1 & 0 & 0 \\
-1 & 0 & 0 & 0 \\
0 & 0 & 0 & -1 \\
0 & 0 & 1 & 0
\end{array} \right | \; .
\label{1.6}
\end{eqnarray}

\noindent Analogously
\begin{eqnarray}
(\alpha^{2})^{2} =
\left | \begin{array}{rrrr}
c_{0}c_{0}+c_{2} & 0 & c_{0} & 0 \\
c_{1}c_{0}+c_{3} & -1 & c_{1} & 0 \\
c_{2}c_{0} & 0 & c_{2} & 0 \\
c_{3}c_{0}-c_{1} & 0 & c_{3} & -1
\end{array} \right | = -I\; ,
\nonumber
\end{eqnarray}

\noindent that is
\begin{eqnarray}
c_{0}=0\; , \; c_{1}= 0 \; , \; c_{2}=-1 \; , \;
c_{3}=0 \;, \qquad
\alpha^{2} = \left | \begin{array}{rrrr}
0 & 0 & 1 & 0 \\
0 & 0 & 0 & 1 \\
-1 & 0 & 0 & 0 \\
0 & -1 & 0 & 0
\end{array} \right | \; , \qquad (\alpha^{2})^{2} = -I\; .
\label{1.7}
\end{eqnarray}

\noindent And finally
\begin{eqnarray}
(\alpha^{3})^{2} =
\left | \begin{array}{rrrr}
d_{0}d_{0}+d_{3} & 0 & 0 & d_{0} \\
d_{1}d_{0}-d_{2} & -1 & 0 & 0 \\
d_{2}d_{0}+d_{1} & 0 & -1 & d_{2} \\
d_{3}d_{0}& 0 & 0 & d_{3}
\end{array} \right | = - I \; ,
\nonumber
\end{eqnarray}

\noindent that is
\begin{eqnarray}
d_{0}=0 \; , \; d_{1}= 0 \; , \; d_{2}=0 \; , \;
d_{3}=-1 \;, \qquad
\alpha^{3}= \left | \begin{array}{rrrr}
0 & 0 & 0 & 1 \\
0 & 0 & -1 & 0 \\
0 & 1 & 0 & 0 \\
-1 & 0 & 0 & 0
\end{array} \right | \; , \qquad (\alpha^{3})^{2} = -I\; .
\label{1.8}
\end{eqnarray}

Consider their products:
\begin{eqnarray}
\alpha^{1} \alpha^{2} = -\alpha^{2} \alpha^{1}  = + \alpha^{3}\;,
\qquad
\alpha^{2} \alpha^{3} =  -\alpha^{3} \alpha^{2}  =
 \alpha^{1} \; , \qquad
 \alpha^{3} \alpha^{1} = - \alpha^{1} \alpha^{3} = \alpha^{2}\; .
\label{1.9c}
\end{eqnarray}

Consider the product
$\alpha^{0}\alpha^{i}$:
\begin{eqnarray}
k=\pm 1 , \;\; \alpha^{0}\alpha^{1}=
\left | \begin{array}{rrrr}
0 & k & 0 & 0 \\
-1 & 0 & 0 & 0 \\
0 & 0 & 0 & -1 \\
0 & 0 & 1 & 0
\end{array} \right | \; , \qquad
 \alpha^{1}\alpha^{0}=  \left | \begin{array}{rrrr}
0 & 1 & 0 & 0 \\
-k & 0 & 0 & 0 \\
0 & 0 & 0 & -1 \\
0 & 0 & 1 & 0
\end{array} \right | \; .
\nonumber
\end{eqnarray}

\noindent If $k=+1$,  we will have the most simple commutation rule:
\begin{eqnarray}
\alpha^{0}= I\; , \qquad \alpha^{i}\alpha^{0} =
\alpha^{0}\alpha^{i}= \alpha^{i} \; . \label{1.9d}
\end{eqnarray}

Thus, the Maxwell matrix equation looks
\begin{eqnarray}
(-i \partial_{0} + \alpha^{j} \partial_{j} ) \Psi =J \; , \qquad
\Psi = \left | \begin{array}{c} 0 \\\psi^{1} \\\psi^{2} \\
\psi^{3}
\end{array} \right | \; , \qquad J=
{1 \over \epsilon_{0}} \; \left | \begin{array}{c} j^{0} \\ i\;
j^{1} \\ i\; j^{2} \\ i \; j^{3}
\end{array} \right | \; ,
\nonumber
\\
\alpha^{1} = \left | \begin{array}{rrrr}
0 & 1 & 0 & 0 \\
-1 & 0 & 0 & 0 \\
0 & 0 & 0 & -1 \\
0 & 0 & 1 & 0
\end{array} \right | \; , \qquad
\alpha^{2} = \left | \begin{array}{rrrr}
0 & 0 & 1 & 0 \\
0 & 0 & 0 & 1 \\
-1 & 0 & 0 & 0 \\
0 & -1 & 0 & 0
\end{array} \right | \; , \qquad
\alpha^{3} = \left | \begin{array}{rrrr}
0 & 0 & 0 & 1 \\
0 & 0 & -1 & 0 \\
0 & 1 & 0 & 0 \\
-1 & 0 & 0 & 0
\end{array} \right |\; ,
\nonumber
\\
(\alpha^{1})^{2} = -I \; , \qquad (\alpha^{1})^{2} = -I \; ,
\qquad (\alpha^{1})^{2} = -I \; , \nonumber
\\
\alpha^{1} \alpha^{2}= - \alpha^{2} \alpha^{1} = \alpha^{3} \;,
\qquad \alpha^{2} \alpha^{3} = - \alpha^{3} \alpha^{2} =
\alpha^{1}\;, \qquad \alpha^{3} \alpha^{1} = - \alpha^{1}
\alpha^{3} = \alpha^{2}\;. \label{1.10}
\end{eqnarray}

\section{ Maxwell matrix equation in Riemannian space}

Maxwell equation
\begin{eqnarray}
(\alpha^{0} \partial_{0} + \alpha^{j} \partial_{j} )\;\Psi =J \; ,
\qquad \alpha^{0} = -i I\; , \nonumber
\\
\Psi = \left | \begin{array}{c} 0 \\ {\bf E} + i c{\bf B}
\end{array} \right | \; , \qquad J
= {1 \over \epsilon_{0}} \; \left | \begin{array}{c} \rho \\
i{\bf j}
\end{array} \right |
\label{2.1}
\end{eqnarray}

\noindent can be extended to an arbitrary Riemannian space-time in
accordance  with general tetrad recipe of
Tetrode-Weyl-Fock-Ivanenko (see  \cite{18}):
\begin{eqnarray}
 \alpha^{\rho}(x)\;[ \; \partial_{\rho } + A_{\rho}(x)\; ] \; \Psi (x) = J(x) \; ,
\nonumber
\\
\alpha^{\rho}(x) = \alpha^{c} \; e_{(c)}^{\rho}(x) \; , \qquad
A_{\rho}(x) = {1 \over 2} j^{ab} \; e_{(a)} ^{\beta} \;
\nabla_{\rho} e_{(n) \beta} \; . \label{2.2}
\end{eqnarray}

\noindent where  $e_{(c)}^{\rho}(x)$ is a tetrad;
$j^{ab}$ stands for generators for complex vector representation of orthogonal
group $SO(3.C)$;  $\nabla_{\rho} $  denotes a covariant derivative.
Eq.  (\ref{2.1}) can be rewritten differently
\begin{eqnarray}
\alpha^{c} \; ( \; e_{(c)}^{\rho} \partial_{\rho} + {1 \over 2}
j^{ab} \gamma_{abc} \; ) \; \Psi = J(x) \; , \label{2.3a}
\end{eqnarray}

\noindent with the use of Ricci  rotation  coefficients$
\gamma_{bac} = - \gamma_{abc} = - e_{(b)\beta ;\alpha}
e^{\beta}_{(a)} e^{\alpha}_{(c)} \;$.

Eq.  (\ref{2.1}) is invariant under gauge  transformations of the
local Lorentz group (see  \cite{18})
 \begin{eqnarray}
 \Psi' (x) = S(x) \Psi (x)\; , \qquad  S(x) \in SO(3.C)_{loc}\; ,
 \nonumber
 \\
  e_{(a)\alpha}' (x) = L_{a}^{\;\;b} (x) \; e_{(b)\alpha} (x) \; ,
\nonumber
\\[3mm]
 \alpha^{\rho}(x)\;[ \partial_{\rho } + A_{\rho}(x) ] \; \Psi (x) = J(x) \; ,
\nonumber
\\
 \alpha^{'\rho}(x)\;[ \partial_{\rho } + A'_{\rho}(x) ] \; \Psi' (x) = J'(x) \; .
\label{2.4}
\end{eqnarray}

\section{ Tetrad explicit form of Maxwell matrix equation
 }

Matrix Maxwell equation (\ref{2.3a}) can be written as
\begin{eqnarray}
-i \; ( \; e_{(0)}^{\rho} \partial_{\rho} + {1 \over 2} j^{ab}
\gamma_{ab0} \; )\Psi + \alpha^{k} \; ( \; e_{(k)}^{\rho}
\partial_{\rho} + {1 \over 2} j^{ab} \gamma_{abk} \; )\Psi =
J(x)\; . \label{3.2}
\end{eqnarray}

\noindent Taking into account the identities
\begin{eqnarray}
{1 \over 2} j^{ab} \gamma_{ab0} = [ s_{1} ( \gamma_{230} +i
\gamma_{010} ) + s_{2} ( \gamma_{310} +i \gamma_{020}) + s_{3} (
\gamma_{120} +i \gamma_{030} )\; ] \; , \nonumber
\\
{1 \over 2} j^{ab} \gamma_{abk} = [ s_{1} ( \gamma_{23k} +i
\gamma_{01k} ) + s_{2} ( \gamma_{31k} +i \gamma_{02k}) + s_{3} (
\gamma_{12k} +i \gamma_{03k} )\; ] \label{3.3}
\end{eqnarray}

\noindent and using notation
\begin{eqnarray}
e_{(0)}^{\rho} \partial_{\rho} = \partial_{(0)} \; , \qquad
e_{(k)}^{\rho} \partial_{\rho} = \partial_{(k)} \; , \qquad \qquad
a =0,1,2,3 \; , \nonumber
\\
( \gamma_{01a}, \gamma_{02a} , \gamma_{03a} ) = {\bf v}_{a} \; ,
\qquad ( \gamma_{23a}, \gamma_{31a} , \gamma_{12a} ) = {\bf p}_{a}
\; , \label{3.4}
\end{eqnarray}

\noindent eq.  (\ref{3.2}) is reduced to
\begin{eqnarray}
-i \; [ \; \partial_{(0)} + {\bf s} ({\bf p}_{0} +i{\bf v}_{0} \;
)\; ] \;\Psi + \alpha^{k} \; [\; \; \partial_{(k)} + {\bf s} ({\bf
p}_{k} +i{\bf v}_{k} \; )\; ] \;\Psi = J(x)\; , \nonumber
\end{eqnarray}

\noindent or
\begin{eqnarray}
(\; \alpha^{k} \; \partial_{(k)} + {\bf s} {\bf v}_{0} +
 \alpha^{k} \; {\bf s} {\bf p}_{k} \; )\; \left | \begin{array}{c}
0 \\ {\bf E} + i c{\bf B}
\end{array} \right |-
\nonumber
\\
-
 i\;
 ( \; \partial_{(0)} +
  {\bf s} {\bf p}_{0} - \alpha^{k} {\bf s} {\bf v}_{k}) \;\left | \begin{array}{c}
0 \\ {\bf E} + i c{\bf B}
\end{array} \right |
 = {1 \over \epsilon_{0}}
 \left | \begin{array}{c}
 \rho \\ i \;{\bf j}
 \end{array} \right | ,
\label{3.5}
\end{eqnarray}

\noindent where  $s_{i}$ stands for the generators
\begin{eqnarray}
s_{1}= \left | \begin{array}{rrrr}
0 & 0 & 0 & 0 \\
0 & 0 & 0 & 0 \\
0 & 0 & 0 & -1 \\
0 & 0 & 1 & 0 \\
\end{array} \right |, \;
s_{2} = \left | \begin{array}{rrrr}
0 & 0 & 0 & 0 \\
0 & 0 & 0 & 1 \\
0 & 0 & 0 & 0 \\
0 & -1 & 0 & 0 \\
\end{array} \right |\; , \;
s_{3} = \left | \begin{array}{rrrr}
0 & 0 & 0 & 0 \\
0 & 0 & -1 & 0 \\
0 & 1 & 0 & 0 \\
0 & 0 & 0 & 0
\end{array} \right | \; .
\nonumber
\end{eqnarray}

\section{ Cylindric coordinates and tetrad in spherical space  $S_{3}$
}

Let us consider the Maxwell equation in the cylindric coordinates and tetrad
in spherical space $S_{3}$:
\begin{eqnarray}
n_{1} = \sin r \; \cos \phi \; , \;\; n_{2} = \sin r \; \sin \phi
\; , \;\;
 n_{3} = \cos r \; \sin z \; , \;\; n_{4} =
\cos r \; \cos z \; ; \nonumber
\\
 d S^{2} = dt^{2} - d r^{2} -
\sin^{2} r \; d\phi^{2} - \cos^{2}r \; dz^{2} \; , \qquad
x^{\alpha} = (t, r , \phi, z)\; , \nonumber
\\[3mm]
 e_{(a)}^{\beta}(y) = \left |
\begin{array}{llll}
1 & 0 & 0 & 0 \\
0 & 1 & 0 & 0 \\
0 & 0 & \sin^{-1} r & 0 \\
0 & 0 & 0 & \cos^{-1}r
\end{array} \right | \; , \;
 e_{(a) \beta}(y) = \left |
\begin{array}{llll}
1 & 0 & 0 & 0 \\
0 & -1 & 0 & 0 \\
0 & 0 & - \sin r & 0 \\
0 & 0 & 0 & - \cos r
\end{array} \right | \; ;
\label{4.1}
\end{eqnarray}

\noindent where $(r,\; \phi,\; z )$ run within
\begin{eqnarray}
  \rho \in [ 0, \; +\pi /2 ] \; , \;\; \phi \in [-\pi,\; +\pi ] \; ,
\;\; z \in [-\pi,\;+\pi ] \; .
\nonumber
\end{eqnarray}

 Christoffel symbols are given by
$ \Gamma^{0}_{\beta \sigma} = 0 \; , \; \Gamma^{i}_{00} = 0 \; ,\;
\Gamma^{i}_{0j} = 0 $ and
\begin{eqnarray}
\Gamma^{r}_{\;\; jk} = \left | \begin{array}{ccc}
0 & 0 & 0 \\
0 & - \sin r \cos r & 0 \\
0 & 0 & \sin r \cos r
\end{array} \right | \; ,
\nonumber
\\
\Gamma^{\phi}_{\;\; jk} = \left | \begin{array}{ccc}
0 & {\cos r \over \sin r} & 0 \\
{\cos r \over \sin r} & 0 & 0 \\
0 & 0 & 0
\end{array} \right | \; , \qquad
\Gamma^{z}_{\;\; jk} = \left | \begin{array}{ccc}
0 & 0 & -{\sin r \over \cos r} \\
0 & 0 & 0 \\
-{\sin r \over \cos r} & 0 & 0
\end{array} \right | \; .
\label{4.4}
\end{eqnarray}

\noindent For covariant derivatives of tetrad vectors we get
\begin{eqnarray}
A_{\beta ; \alpha }= {\partial A_{\beta} \over \partial
x^{\alpha}} - \Gamma^{\sigma} _{\alpha \beta} A_{\sigma} \qquad
\Longrightarrow \qquad
e_{(0) \beta ; \alpha }= {\partial e_{(0)\beta} \over \partial x^{\alpha}} - \Gamma^{0}
_{\alpha \beta} \; e_{(0) 0} = 0 \; ,
\nonumber
\end{eqnarray}

\begin{eqnarray}
e_{(1) \beta ; \alpha }=
\Gamma^{r} _{\alpha \beta} \qquad  \Longrightarrow \qquad
e_{(1) \beta ; \alpha } = \left | \begin{array}{cccc}
0 & 0 & 0 & 0 \\
0 & 0 & 0 \\
0 & 0 & - \sin r \cos r & 0 \\
0 & 0 & 0 & \sin r \cos r
\end{array} \right | \; ,
\nonumber
\end{eqnarray}

\begin{eqnarray}
e_{(2) \beta ; \alpha }= {\partial e_{(2)\beta} \over \partial
x^{\alpha}} - \Gamma^{\phi} _{\alpha \beta} \; e_{(2)\phi}   =   \left | \begin{array}{cccc}
0 & 0 & 0 & 0 \\
0 & 0 & \cos r & 0 \\
0 & 0 & 0 & 0 \\
0 & 0 & 0 & 0
\end{array} \right | ,
\nonumber
\end{eqnarray}

\begin{eqnarray}
e_{(3) \beta ; \alpha }= {\partial e_{(3)\beta} \over \partial
x^{\alpha}} - \Gamma^{z} _{\alpha \beta} \; e_{(3) z } =    \left | \begin{array}{cccc}
0 & 0 & 0 & 0 \\
0 & 0 & 0 & -\sin r \\
0 & 0 & 0 & 0 \\
0 & 0 & 0 & 0
\end{array} \right | .
\nonumber
\end{eqnarray}

It remains to find Ricci rotation coefficients:
\begin{eqnarray}
\gamma_{ab 0} = e_{(a)}^{\;\;\; \beta} \; e_{(b)\beta; t } \;
e_{(0)}^{t} = 0 \; ,  \qquad
\gamma_{ab 1} = e_{(a)}^{\;\; \; a } \; e_{(b)a; r } \;
e_{(1)}^{r} \; ,
\nonumber
\\
\gamma_{ab 2} = e_{(a)}^{\;\; \; \beta} \;
e_{(b)\beta; \phi} \; e_{(2)}^{\phi}\; , \qquad \gamma_{ab 3} =
e_{(a)}^{\;\; \; \beta} \; e_{(b)\beta; z } \; e_{(3)}^{z} \; ;
\nonumber
\end{eqnarray}

\noindent from which it follow
\begin{eqnarray}
\gamma_{01 1}= \gamma_{02 1}= \gamma_{03 1} = 0 \; , \qquad
\gamma_{012}= \gamma_{022}= \gamma_{032} = 0 \; , \qquad
\gamma_{01 3}= \gamma_{02 3}= \gamma_{03 3} = 0 \; ,
\nonumber
\end{eqnarray}
\begin{eqnarray}
\gamma_{23 1} =
0 \; , \qquad  \gamma_{31 1} =  0 \; , \qquad  \gamma_{12 1} = 0 \; , \nonumber
\\
\gamma_{23 2} = 0 \; , \;\; \gamma_{31 2} =  0 \; , \;\; \gamma_{12 2}
=
 { \cos r \over \sin r} \; ,
\nonumber
\\
\gamma_{23 3} =  0 \; , \qquad  \gamma_{31 3} =
   {\sin r \over \cos r }\; ,\qquad
\gamma_{12 3} = 0 \; . \nonumber
\end{eqnarray}

Taking into account the identities
\begin{eqnarray}
e_{(0)}^{\rho} \partial_{\rho} = \partial_{(0)} = \partial_{t} \;
, \qquad e_{(1)}^{\rho} \partial_{\rho} = \partial_{(1)} =
\partial_{r}\; ,
\nonumber
\\
e_{(2)}^{\rho} \partial_{\rho} = \partial_{(2)} = {1 \over \sin r}
\partial_{\phi} \; , \qquad e_{(3)}^{\rho} \partial_{\rho} =
\partial_{(3)} = {1 \over \cos r} \partial_{z} \; ,
\nonumber
\\
 {\bf v}_{0} =( \gamma_{01 0}, \gamma_{02 0} , \gamma_{03 0} ) \equiv 0 \; , \qquad
 {\bf v}_{1} = ( \gamma_{01 1}, \gamma_{021 } , \gamma_{03 1} ) \equiv 0 \; ,
\nonumber
\\
 {\bf v}_{2} = ( \gamma_{012 0}, \gamma_{02 2} , \gamma_{03 2} ) \equiv 0 \; ,
\qquad
 {\bf v}_{3} = ( \gamma_{013}, \gamma_{02 3} , \gamma_{03 3} ) \equiv 0 \; ,
\nonumber
\\
 {\bf p}_{0} = ( \gamma_{23 0}, \gamma_{31 0} , \gamma_{12 0} ) = 0 \; , \qquad
 {\bf p}_{1} = ( \gamma_{23 1}, \gamma_{31 1} , \gamma_{12 1} ) = 0 \; ,
\nonumber
\\
 {\bf p}_{2} = ( \gamma_{23 2}, \gamma_{31 2} , \gamma_{12 2} ) = (0 , 0, {\cos r \over \sin r}) \; ,
\qquad
 {\bf p}_{3} = ( \gamma_{23 3}, \gamma_{31 3} , \gamma_{12 3} ) = (0, {\sin r \over \cos r} , 0 ) \; ,
\nonumber
\\
\label{4.10}
\end{eqnarray}

\noindent in the absence of an external source eq. (\ref{3.5})  reads
\begin{eqnarray}
\left (\; - i \partial_{t} + \alpha^{1} \; \partial_{r} +
\alpha^{2} \; {1 \over \sin r} \partial_{\phi} +
 \alpha^{3} \;{1 \over \cos r}\partial_{z}
 + \alpha^{2} \; S_{3} \; {\cos r \over \sin r} +\alpha^{3} \; S_{2} \; {\sin r \over \cos r} \; \right )
 \left | \begin{array}{c}
0 \\ {\bf E} + i c{\bf B}
\end{array} \right | = 0 \; .
\nonumber
\\
\label{4.11}
\end{eqnarray}

\section{ Separation of variables  in $S_{3}$, solutions at $m=0$ }

Wave Maxwell operator from (\ref{4.11}) commutes with the following ones:
 $i\partial_{t},\; i\partial_{\phi},\;i\partial_{z}$. Therefore, for a field function
 we get  a substitution
 \begin{eqnarray}
\Psi = \left | \begin{array}{c} 0 \\ {\bf E} + i c{\bf B}
\end{array} \right | =
 e^{- i \omega t} \; e^{im\phi} \; e^{ikz} \;
 \left |
\begin{array}{c} 0 \\ f_{1}(r) \\ f_{2}(r) \\ f_{3}(r)
\end{array} \right | \; .
\label{5.2}
\end{eqnarray}

\noindent Correspondingly, eq. (\ref{4.11}) reads
\begin{eqnarray}
\left (\; - \omega + \alpha^{1} \; {d \over dr} + {im \over \sin
r} \; \alpha^{2} + { ik \over \cos r} \; \alpha^{3} + {\cos r
\over \sin r} \; \alpha^{2} S_{3} + {\sin r \over \cos r}\;
\alpha^{3} S_{2} \; \right ) \left |
\begin{array}{c} 0 \\ f_{1}(r) \\ f_{2}(r) \\ f_{3}(r)
\end{array} \right |
 = 0 \; .
\label{5.3}
\end{eqnarray}

\noindent After simple calculation we get the following radial system:
\begin{eqnarray}
({d \over dr } + {\cos r \over \sin r} - {\sin r \over \cos r })
f_{1} +
 {im \over \sin r}\; f_{2} + {ik \over \cos r }\; f_{3} = 0 \; ,
\nonumber
\\
-\omega f_{1} - {ik \over \cos r}\; f_{2} + {im \over \sin r}\;
f_{3} = 0 \; , \nonumber
\\
-\omega f_{2} - ({d \over dr} - {\sin r \over \cos r} )\; f_{3} +
{ik \over \cos r} f_{1} = 0 \; , \nonumber
\\
-\omega f_{3} + ({d \over dr} + {\cos r \over \sin r} )\; f_{2} -
{im \over \sin r} f_{1} = 0 \; . \label{5.5}
\end{eqnarray}

Let us consider first the case $m=0$, when the radial equations  become  more simple
\begin{eqnarray}
({d \over dr } + {\cos r \over \sin r} - {\sin r \over \cos r })
f_{1}
 + {ik \over \cos r }\; f_{3} = 0 \; ,
\nonumber
\\
 f_{1} = -{ik \over \omega \; \cos r}\; f_{2} \; ,
\nonumber
\\
-\omega f_{2} - ({d \over dr} - {\sin r \over \cos r} )\; f_{3} +
{ik \over \cos r} f_{1} = 0 \; , \nonumber
\\
f_{3} = {1 \over \omega}\; ({d \over dr} + {\cos r \over \sin r}
)\; f_{2} \; . \label{5.7}
\end{eqnarray}

Using 2nd and 4th equations, from the first one it follows
\begin{eqnarray}
({d \over dr } + {\cos r \over \sin r} - {\sin r \over \cos r })
\; {-ik \over \omega \; \cos r}\; f_{2}
 + {ik \over \cos r }\; {1 \over \omega}\; ({d \over dr} + {\cos r \over \sin r} )\; f_{2} = 0 \; ,
\nonumber
\\
-\omega f_{2} - ({d \over dr} - {\sin r \over \cos r} )\; {1 \over
\omega}\; ({d \over dr} + {\cos r \over \sin r} )\; f_{2}
  + {ik \over \cos r} \; {-ik \over \omega \; \cos r}\; f_{2} = 0 \; ,
\nonumber
\end{eqnarray}

\noindent that is equivalent to the identity
 $0 \equiv 0$ and the equation for
$f_{2}$:
\begin{eqnarray}
{d^{2} \over dr^{2}} \; f_{2} + ( {\cos r \over \sin r } - {\sin r
\over \cos r} ) \; {d \over dr} \; f_{2} + ( \omega^{2} -1 - {1
\over \sin^{2}r} - {k^{2} \over \cos^{2} r} )\; f_{2} = 0 \; .
\label{5.8b}
\end{eqnarray}

\noindent
The latter  can be simplified:
\begin{eqnarray}
f_{2}(r) = {1 \over \sin r}\; E (r) \; , \qquad
{d^{2} E \over dr^{2}} - {1 \over \sin r \cos r}\; {dE \over dr} +
( \omega^{2} - {k^{2} \over \cos^{2} r} ) E = 0\; ;
 \label{5.10a}
\end{eqnarray}

\noindent two  concomitant functions are given by
\begin{eqnarray}
 f_{1} (r) = {-ik \over \omega } \; {1 \over \cos r \sin r }\; E(r) \; , \qquad
f_{3} = {1 \over \omega}\; {1 \over \sin r} \; {d \over dr }E(r)
\; .
\nonumber
\label{5.10b}
\end{eqnarray}

Turning back to eq. (\ref{5.10a}),  first let us consider  a particular case when $k^{2}=
\omega^{2}$:
\begin{eqnarray}
 {\sin r \over \cos r} {d \over dr} { \cos r \over \sin r} {d \over dr} \; E
    + k^{2} (1 - {1 \over \cos^{2} r} ) \; E = 0 \; ;
\nonumber
\end{eqnarray}

\noindent from whence it follows
\begin{eqnarray}
( { \cos r \over \sin r} {d \over dr} )\;
 ( { \cos r \over \sin r} {d \over dr} ) \; E = k^{2}\; E \; ,
\nonumber
\label{6.2}
\end{eqnarray}

\noindent so that
\begin{eqnarray}
{ \cos r \over \sin r } { d \over dr } = {d \over dx } \; , \qquad
\Longrightarrow \qquad {d r \over dx } = { \cos r \over \sin r }
\; , \nonumber
\\
 -dx = d \log \cos r \;, \qquad x =
  \log ( C \; \cos^{-1} r ) \; , \qquad C = \mbox{const}
\nonumber
\label{6.3a}
\end{eqnarray}

\noindent and eq.  (\ref{6.2}) reads
\begin{eqnarray}
{d^{2} \over dx^{2} } \; E = k^{2} \; E \; , \nonumber
\end{eqnarray}

\noindent which has two solutions
\begin{eqnarray}
k^{2}=\omega^{2}\; , \qquad  E_{\pm} = e^{\mp k x } = E_{0} \; ( \cos r )^{\pm k}\; .
\label{6.3b}
\end{eqnarray}

\noindent Among solutions
\begin{eqnarray}
( \cos r )^{ + k}\; , \qquad {1 \over ( \cos r )^{ k} } \; ,
\qquad r \in [\; 0 , {\pi \over 2 } \; ] \label{6.4}
\end{eqnarray}

\noindent at $k>0$  the second one  must be rejected because it turns to infinity  as
 $r \rightarrow  \pi /2 $;  when  for $k<0$ the first one must be rejected  by analogous reason.
Thus, physical solutions are
\begin{eqnarray}
k^{2}=\omega^{2}\; , \; k>0\;, \qquad E = E_{0} \; \cos ^{ k} r \; e^{-i (\omega t - kz )}
\; ;\;
 \nonumber
\\[2mm]
k^{2}=\omega^{2}\; , \; k<0\;, \qquad E = E_{0} \; \cos ^{ -k} r \; e^{-i (\omega t - kz
)} \; . \label{6.5}
\end{eqnarray}

Because at  $r \neq 0 , \pi /2$, the values  $z = - \pi$
and $z= + \pi $ determine one the same point in spherical space $S_{3}$,  solutions
(\ref{6.5})  represent continuous  function in $S_{3}$ only if $k$ takes on integer values:
\begin{eqnarray}
k = \pm n \; , \qquad n = 1,2, 3, ... \; ; \label{6.6a}
\end{eqnarray}

\noindent or in usual units
\begin{eqnarray}
k = \pm {\omega \rho \over c } = n \; , \qquad \omega = {c \over
\rho} \; n \; , \qquad n = 1, 2, 3, ... \label{6.6b}
\end{eqnarray}

Turning again to  eq.  (\ref{5.10a}) let us construct  one more special solution.
To this end, considering approximate solution in the vicinity of the point $r=0$:
$
E = \sin^{A} r $, we get
$
 A = 0, \; +2$.
In the same manner nearby the point $r=\pi /2 $ an approximate solution is
$
E = \cos^{B} r\; ,\;  B ^{2} = k^{2}$.
Let us demonstrate that there exist values  $k$ such that an exact solution can be constructed as follows
(the case $B=0$ was  considered above):
\begin{eqnarray}
E = \sin^{2} r \cos^{B}r \;  . \label{6.9a}
\end{eqnarray}

\noindent  With this substitution,  eq. (\ref{5.10a}) gives
\begin{eqnarray}
2 \cos^{4}r - 2(B+1) \sin^{2} r\; \cos^{2}r -3B\sin^{2}r \;
\cos^{2}r + B(B-1) \sin^{4}r \nonumber
\\
- 2 \cos^{2} r + B \sin^{2} r \; - k^{2} \sin^{2} r + \omega^{2}
\sin^{2} r \cos^{2}r =0 \; . \nonumber
\end{eqnarray}

\noindent With notation  $ \cos^{2} r = x  $ it can be written as
\begin{eqnarray}
2 \; x^{2} + (x -x^{2}) \; [ - 5B -2 + \omega^{2} ] + (B^{2} - B)
(1-2x +x^{2}) - 2 x + B (1-x) - k^{2} (1-x) =0 \;, \nonumber
\end{eqnarray}

\noindent or
\begin{eqnarray}
x^{2} \; ( 4 + 4B - \omega^{2} + B^{2} ) +
  x \; ( -4 - 4B + \omega ^{2} - B^{2} ) +
x^{0} \; ( B^{2} -k^{2} ) = 0 \; . \nonumber
\end{eqnarray}

\noindent The latter  is satisfied if
\begin{eqnarray}
B^{2} = k^{2} \; , \qquad (B +2)^{2} - \omega^{2} = 0 \; ,
\nonumber
\end{eqnarray}

\noindent that is
\begin{eqnarray}
B = -2 + \omega, \; -2 - \omega \; ,
 \nonumber
\\
 k = \pm B
\;, \qquad
 E = \sin^{2} r \; \; \cos^{ B}r \; .
\label{6.10b}
\end{eqnarray}

Thus, the corresponding solutions of this type are
\begin{eqnarray}
E = \sin^{2} r \; \; \cos^{B}r \; e^{-i (\omega t - k z)}\; .
\label{6.10c}
\end{eqnarray}

\noindent Solutions with  negative  $B$ must be rejected because they give infinite  electromagnetic field
at the point
 $r = \pi /2$. Besides, periodicity requirement   on $z$ leads to
 $k = \pm 1, \pm 2, \pm 3, ...$.

Therefore, the wave propagating  in the positive direction is given by
\begin{eqnarray}
B = + k = +1, +2, +3, ... , \nonumber
\\
k = -2 + \omega \; , \qquad \omega = 2 + k = 3,4, 5, ... \; ,
\nonumber
\\
E = \sin^{2} r \; \; \cos^{k}r \; e^{-i (\omega t - k z)} \;
. \label{6.11a}
\end{eqnarray}

In turn, the wave propagating  in the negative direction  is given by
\begin{eqnarray}
B = -k = +1, +2, +3, ... , \nonumber
\\
-k = -2 + \omega \; , \qquad \omega = 2 - k = +3,+4, +5, ... \; ,
\nonumber
\\
E = \sin^{2} r \; \; \cos^{-k}r \; e^{-i (\omega t - k z) }
\; . \label{6.11b}
\end{eqnarray}

 Turning to the general equation  (\ref{5.10a}), one may try to construct
 all other solutions of that type $m=0$ on the base of the following substitution:
 \begin{eqnarray}
E (r) = \sin^{2} r \; \cos^{B}r \; F (r)\; ; \label{6.12b}
\end{eqnarray}

\noindent  eq. (\ref{5.10a}) gives (let  $ \cos^{2} r = x $)
\begin{eqnarray}
4x (1-x) {d^{2} \over dx^{2}} F + 4[ 1 -3x +B(1-x)]\; {d \over d
x} + \nonumber
\\
+ \left [ -(2 + 5 B) + {2x \over 1-x} +B(B-1) {1-x \over x} - {2
\over 1-x} +{B \over x}-{k^{2} \over x} + \omega^{2} \right ] F=0
\; .
\nonumber
\label{6.12c}
\end{eqnarray}

\noindent Requiring $ k^{2}=B^{2} $, for  $F$ we get the equation
\begin{eqnarray}
4x (1-x) {d^{2} \over dx^{2}} F + 4[ (1+B)- (3+B)x ]\; {d \over d
x} - [\; (B+2)^{2} - \omega^{2} \; ]\; F =0\; ,
\nonumber
\label{6.13b}
\end{eqnarray}

\noindent which is of  hypergeometric  type
\begin{eqnarray}
z(1-z) \; F + [ \gamma - (\alpha + \beta +1) z ] \; F' - \alpha
\beta \; F = 0 \; ,
\nonumber
\\
k= \pm \; B \; , \qquad \gamma = 1 + B\; , \;\; \alpha = {B+2 -
\omega \over 2} \;, \;\; \beta = { B+2 + \omega \over 2} \; .
\nonumber
\label{6.13c}
\end{eqnarray}

\noindent Thus, the general solution of the type $m=0$ takes the form
\begin{eqnarray}
E = \sin^{2} r \cos^{B}r \; F(\alpha, \beta, \gamma, \cos^{2} r
)\; e^{-i (\omega t - k z) }\; . \label{6.13d}
\end{eqnarray}

\noindent
We are to  separate single-valued and continuous functions in $S_{3}$.

\vspace{2mm} For a wave propagating in the positive direction  $z$:
\begin{eqnarray}
k > 0\; , \qquad k = +1, +2, +3 , ... \; ; \label{6.14a}
\end{eqnarray}

\noindent the function  $E(r)$ is finite at $r = \pi/2$ only if
$B =+ k$, besides polynomial  solutions arise only if
\begin{eqnarray}
\alpha = {k+2 - \omega \over 2} = -n\; = 0, -1, -2, ...\qquad
\Longrightarrow \nonumber
\\
 \omega = k +2(n+1)= N\; .
\label{6.14b}
\end{eqnarray}

\vspace{2mm} For a wave propagating in the positive direction  $z$:
\begin{eqnarray}
k<0\; , \qquad -k = 1, 2, 3 , ... \; ; \label{6.15a}
\end{eqnarray}

\noindent the function  $E(r)$ is finite at  $r = \pi/2$ only if
$B =- k$, additionally  one must obtain polynomials which leads to
\begin{eqnarray}
\alpha = {-k+2 - \omega \over 2} = -n\; = 0, -1, -2, ...\qquad
\Longrightarrow \nonumber
\\
\omega = -k +2(n+1)= N\; . \label{6.15b}
\end{eqnarray}

All constructed solutions of the Maxwell equations are finite, single-valued, and continuous  functions
in  spherical Riemann space  $S_{3}$.

\section{Maxwell solutions at  $k=0$}

Turning to eqs.
(\ref{5.5}) at  $k=0$:
\begin{eqnarray}
({d \over dr } + {\cos r \over \sin r} - {\sin r \over \cos r })
f_{1} +
 {im \over \sin r}\; f_{2} = 0 \; ,
\nonumber
\\
f_{1} ={im \over \omega \sin r}\; f_{3} \; , \nonumber
\\
 f_{2} =-{1 \over \omega} ({d \over dr} - {\sin r \over
\cos r} )\; f_{3} \; , \nonumber
\\
-\omega f_{3} + ({d \over dr} + {\cos r \over \sin r} )\; f_{2} -
{im \over \sin r} f_{1} = 0 \; . \label{7.3}
\end{eqnarray}

With the use of second and  third from  first and fourth we get
 an identity
 $0 \equiv 0$ and the following equation for
$f_{3}$:
\begin{eqnarray}
{d^{2} \over dr^{2}} \; f_{3} + ( {\cos r \over \sin r } - {\sin r
\over \cos r} ) \; {d \over dr} \; f_{3} + ( \omega^{2} -1 - {1
\over \cos^{2}r} - {k^{2} \over \sin^{2} r} )\; f_{3} = 0 \; .
\label{7.5}
\end{eqnarray}

\noindent which gives
\begin{eqnarray}
f_{3}(r) = {1 \over \cos r}\; E (r) \; , \qquad
{d^{2} E \over dr^{2}} + {1 \over \sin r \cos r}\; {dE \over dr} +
( \omega^{2} - {m^{2} \over \sin^{2} r} ) E = 0\; . \label{7.7}
\end{eqnarray}

First, consider a particular case  $m^{2}= \omega^{2}$:
\begin{eqnarray}
 {d^{2} E \over dr^{2}} + {1 \over \sin r \cos r}\; {dE \over dr}
    + m^{2} (1 - {1 \over \sin^{2} r} ) \; E = 0 \; ;
\label{7.8}
\end{eqnarray}

\noindent
 with the solutions
\begin{eqnarray}
( \sin r )^{  m}\; , \qquad {1 \over ( \sin r )^{ m} } \; ,
\qquad r \in [\; 0 , {\pi \over 2 } \; ] \; . \label{7.10}
\end{eqnarray}

\noindent Physical solutions are
\begin{eqnarray}
m^{2}= \omega^{2}\; , \; m>0\;, \qquad E = E_{0} \; \sin ^{ m} r \; e^{-i (\omega t - m
\phi)} \; ;\;
 \nonumber
\\[2mm]
m^{2}= \omega^{2}\; , \; m<0\;, \qquad E = E_{0} \; \sin ^{ -m} r \; e^{-i (\omega t -
m\phi )} \; . \label{7.11}
\end{eqnarray}

Performing analysis like in previous Section,  we easily construct
solutions:
\begin{eqnarray}
B = + m = +1, +2, +3, ... , \nonumber
\\
m = -2 + \omega \; , \qquad \omega = 2 + m = 3,4, 5, ... \; ,
\nonumber
\\
F_{02} = \cos^{2} r \; \; \sin^{m}r \; e^{-i (\omega t - m \phi)}
\; . \label{7.19}
\end{eqnarray}

\noindent and
\begin{eqnarray}
B = -m = +1, +2, +3, ... , \nonumber
\\
-m = -2 + \omega \; , \qquad \omega = 2 - m = +3,+4, +5, ... \; ,
\nonumber
\\
F_{02} = \cos^{2} r \; \; \sin^{-m}r \; e^{-i (\omega t - m z) }
\; . \label{7.20}
\end{eqnarray}

All possible solutions of eq. (\ref{7.7})
can be constructed on the base of a substitution:
\begin{eqnarray}
E (r) = \cos^{2} r \; \sin^{B}r \; F (r)\; . \label{7.22}
\end{eqnarray}

\noindent  and further (let   $\sin^{2} r = x$)
we get
\begin{eqnarray}
m^{2}=B^{2} \;,  \qquad
4x (1-x) {d^{2} \over dx^{2}} F + 4[ (1+B)- (3+B)x ]\; {d \over d
x} - [\; (B+2)^{2} - \omega^{2} \; ]\; F =0\; ,
\nonumber
 \label{7.25}
\end{eqnarray}

\noindent what is of hypergeometric type
\begin{eqnarray}
z(1-z) \; F + [ \gamma - (\alpha + \beta +1) z ] \; F' - \alpha
\beta \; F = 0 \; ,
\nonumber
\\
k= \pm \; B \; , \qquad \gamma = 1 + B\; , \;\; \alpha = {B+2 -
\omega \over 2} \;, \;\; \beta = { B+2 + \omega \over 2} \; .
\label{7.26}
\end{eqnarray}

Thus, the Maxwell equations solutions of the type $k=0$ is given by
\begin{eqnarray}
E = \cos^{2} r \sin^{B}r \; F(\alpha, \beta, \gamma, \sin^{2} r
)\; e^{-i (\omega t - m\phi) }\; . \label{7.27}
\end{eqnarray}

\noindent
We are to separate physical waves.

\vspace{2mm}
$m > 0\; , \qquad B =+ m \;$,
\begin{eqnarray}
\alpha = {m+2 - \omega \over 2} = -n\; \qquad
\Longrightarrow \qquad
 \omega = m +2(n+1)= N\; .
\label{7.29}
\end{eqnarray}

$ m<0\; ,  \qquad B =- m$,
\begin{eqnarray}
\alpha = {-m+2 - \omega \over 2} = -n\; \qquad
\Longrightarrow \qquad
\omega = -m +2(n+1)= N\; . \label{7.31}
\end{eqnarray}

\section{ Radial system at arbitrary $m,k$, general solutions }

Now let us solve radial equations in general case (\ref{5.5}).
The first equation reduces to the identity $0=0$ when taking into account  three remaining:
\begin{eqnarray}
-\omega f_{1} = {ik \over \cos r}\; f_{2} - {im \over \sin r}\;
f_{3} \; , \nonumber
\\
-\omega f_{2} = ({d \over dr} - {\sin r \over \cos r} )\; f_{3} -
{ik \over \cos r} f_{1} \; , \nonumber
\\
-\omega f_{3} = - ({d \over dr} + {\cos r \over \sin r} )\; f_{2}
+ {im \over \sin r} f_{1} \; ;
 \label{5.12}
\end{eqnarray}

\noindent and the first equation takes the form
\begin{eqnarray}
({d \over dr } + {\cos r \over \sin r} - {\sin r \over \cos r })
({ik \over \cos r}\; f_{2} - {im \over \sin r}\; f_{3} ) +
 {im \over \sin r}\; \left [ ({d \over dr} - {\sin r \over \cos r} )\; f_{3} - {ik \over \cos r} f_{1}\right ] +
 \nonumber
\\
 +
  {ik \over \cos r }\; \left [ - ({d \over dr} + {\cos r \over \sin r} )\; f_{2}
   + {im \over \sin r} f_{1} \right ] = 0 \; .
\nonumber
\end{eqnarray}

\noindent what is the identity $0=0$.
The system (\ref{5.12}) is simplified:
\begin{eqnarray}
f_{2} = {1 \over \sin r} \; F_{2}\; , \qquad f_{3} = {1 \over \cos
r} \; F_{3}\; ,
\nonumber
\end{eqnarray}

\noindent so that
\begin{eqnarray}
-\omega \; f_{1} = i \; { k \; F_{2} -
 m \; F_{3} \over \sin r \cos r } \; ,
\nonumber
\\
-\omega \; {F _{2} \over \sin r} = {1 \over \cos r} \; {d F_{3}
\over dr} - {ik \over \cos r} \; f_{1} \; , \nonumber
\\
-\omega \; {F_{3} \over \cos r } = - {1 \over \sin r} \; {d F_{2}
\over dr} + {im \over \sin r}\; f_{1} \; . \label{5.14}
\end{eqnarray}

\noindent Excluding   $f_{1}$, we arrive at
\begin{eqnarray}
  ( {\omega \over \cos r} \; {d \over dr}
     + {k m \over \sin r \cos^{2} r }
   ) \; F_{3}
    + {1 \over \sin r}\; ( \omega^{2} - {k^{2} \over \cos^{2} r } )\; F _{2} = 0 \; ,
\nonumber
\\
 ( {\omega \over \sin r} \; {d \over dr} -
{km \over \cos r \sin^{2} r} ) \; F_{2} + {1 \over \cos r}\; ( -
\omega ^{2}
 + { m ^{2} \over \sin^{2} r}
      ) \; F_{3} = 0 \; .
\label{5.18}
\end{eqnarray}

\noindent
With the use of a new variable  $ y = (1- \cos 2r)/2$ the system reads
\begin{eqnarray}
(2 \omega {d \over dy} - {km \over y (1 - y) }) \; F_{2} +
 ( - { \omega^{2} \over1 - y} + {m^{2} \over y (1-y)})\; F_{3}
= 0\; , \nonumber
\\
(2 \omega {d \over dy} + {km \over y (1 - y) })\; F_{3} + ( + {
\omega^{2} \over y} - {k^{2} \over y (1-y)})\; F_{2} = 0 \; .
\label{8.4}
\end{eqnarray}

Instead of $F_{2},F_{3}$ let us introduce new  functions by means of linear transformation
with unit determinant $\alpha N - \beta M = 1$:
\begin{eqnarray}
F_{2} = \alpha (y) \; G_{2} + \beta (y) \; G_{3} \; , \nonumber
\\
F_{3} = M (y) \; G_{2} + N (y) \; G_{3} \; , \label{8.5}
\end{eqnarray}

\noindent and inverse given by
\begin{eqnarray}
G_{2} = N (y) \; F_{2} - \beta (y) \; F_{3} \; , \nonumber
\\
G_{3} = - M (y) \; F_{2} + \alpha (y) \; F_{3}\; . \label{8.6}
\end{eqnarray}

Combining eqs.  (\ref{8.4}) we get
\begin{eqnarray}
N \; (2 \omega {d \over dy} - {km \over y (1 - y) }) \; F_{2} +
2\omega {d N \over d y} F_{2} - 2\omega {d N \over d y} F_{2} +
 N \; ( - { \omega^{2} \over1 - y} + {m^{2} \over y (1-y)})\; F_{3} -
\nonumber
\\
- \beta \; (2 \omega {d \over dy} + {km \over y (1 - y) })\; F_{3}
- 2\omega {d \beta \over d y} F_{3} + 2\omega {d \beta \over d y}
F_{3 } - \beta \; ( + { \omega^{2} \over y} - {k^{2} \over y
(1-y)})\; F_{2} = 0 \; , \nonumber
\\[3mm]
-M \; (2 \omega {d \over dy} - {km \over y (1 - y) }) \; F_{2} -
2\omega {d M \over d y} F_{2} + 2\omega {d M \over d y} F_{2} - M
( - { \omega^{2} \over1 - y} + {m^{2} \over y (1-y)})\; F_{3} +
\nonumber
\\
 +
\alpha \; (2 \omega {d \over dy} + {km \over y (1 - y) })\; F_{3}
+ 2\omega {d \alpha \over d y} F_{3} - 2\omega {d \alpha \over d
y} F_{3 } + \alpha \; ( + { \omega^{2} \over y} - {k^{2} \over y
(1-y)})\; F_{2} = 0 \; , \label{8.7}
\end{eqnarray}

\noindent from  whence it follows that
\begin{eqnarray}
 2 \omega {d \over dy} \; G_{2} - N \; {km \over y (1 - y) }\; F_{2}
 - 2\omega {d N \over d y} F_{2} +
 N \; ( - { \omega^{2} \over1 - y} + {m^{2} \over y (1-y)})\; F_{3} -
\nonumber
\\
 - \beta \; {km \over y (1 - y) } \; F_{3} + 2\omega {d \beta \over d y} F_{3 } -
\beta \; ( + { \omega^{2} \over y} - {k^{2} \over y (1-y)})\;
F_{2} = 0 \; , \nonumber
\\[3mm]
2 \omega {d \over dy} \; G_{3} + M\; {km \over y (1 - y) } \;
F_{2}
 + 2\omega {d M \over d y} F_{2} -
M \; ( - { \omega^{2} \over1 - y} + {m^{2} \over y (1-y)})\; F_{3}
+ \nonumber
\\
+ \alpha \; {km \over y (1 - y) }\; F_{3} - 2\omega {d \alpha
\over d y} F_{3 } + \alpha ( + { \omega^{2} \over y} - {k^{2}
\over y (1-y)})\; F_{2} = 0 \; . \nonumber
\\
\label{8.8}
\end{eqnarray}

Instead  of $F_{2},F_{3}$ we substitute their expression through  $G_{2}, G_{3}$
according to (\ref{8.5}):
\begin{eqnarray}
2 \omega {dG_{2} \over dy} + \left [ - ( N \alpha + \beta M) \;
{km \over y(1-y)} - 2 \omega {dN \over dy} \; \alpha + NM \; { -
\omega^{2} y + m^{2} \over y(1 - y)} + \right. \nonumber
\\
\left. + 2\omega {d \beta \over dy} M - \beta \alpha\; {
\omega^{2}(1-y) - k^{2} \over y(1-y) } \; \right ] \; G_{2} +
\nonumber
\\[4mm]
+ \left [ - 2 N \beta \; {km \over y(1-y)} - 2 \omega {dN \over
dy} \; \beta + N^{2} {-\omega^{2} y + m^{2} \over y(1 - y) } +
\right. \nonumber
\\
\left. + 2\omega {d \beta \over dy} N - \beta ^{2} \;
{\omega^{2}(1-y) - k^{2} \over y(1-y)} \; \right ] \; G_{3} = 0 \;
, \label{8.11}
\end{eqnarray}

\begin{eqnarray}
2 \omega {dG_{3} \over dy} + \left [
  ( M \beta + \alpha N ) \; {km \over y(1-y)} + 2 \omega {d M \over dy} \; \beta -
NM \; { -\omega^{2}y + m^{2} \over y(1 - y)} - \right. \nonumber
\\
\left. - 2\omega {d \alpha \over dy} N + \beta \alpha\;
{\omega^{2}(1-y) - k^{2} \over y(1-y)} \; \right ] \; G_{3} +
\nonumber
\\[4mm]
+ \left [ 2 M \alpha \; {km \over y(1-y)} + 2 \omega {dM \over dy}
\; \alpha - M^{2} \; {- \omega^{2} y + m^{2} \over y(1 - y) } -
\right. \nonumber
\\
\left. - 2\omega {d \alpha \over dy} M + \alpha ^{2} \;
{\omega^{2}(1-y) - k^{2} \over y(1-y) } \; \right ] \; G_{2} = 0
\; , \label{8.12}
\end{eqnarray}

Let us assume that the transformation used is an orthogonal one:
\begin{eqnarray}
  \alpha \; G_{2} + \beta \; G_{3} = \cos A \; G_{2} + \sin A \; G_{3} \; ,
\nonumber
\\[2mm]
 M \; G_{2} + N \; G_{3} = - \sin A \; G_{2} + \cos A \; G_{3} \; ,
\label{8.13}
\end{eqnarray}

\noindent then
\begin{eqnarray}
- 2 \omega {dN \over dy} \; \alpha + 2\omega {d \beta \over dy} M
= -2 \omega [ ( \cos A)' \cos A + (\sin A)' \sin A ] = 0 \; ,
\nonumber
\\
- 2 \omega {dN \over dy} \; \beta + 2\omega {d \beta \over dy} N =
2 \omega [\; - (\cos A )' \sin A + (\sin A)' \cos A = +\; 2 \omega
\; A ' \; , \nonumber
\\
2 \omega {d M \over dy} \; \beta - 2\omega {d \alpha \over dy} N =
2 \omega [ - (\sin A)' \sin A - (\cos A)' \cos A ] = 0 \; ,
\nonumber
\\
2 \omega {dM \over dy} \; \alpha - 2\omega {d \alpha \over dy} M =
2 \omega [ - (\sin A )' \cos A + (\cos A)' \sin A ] = -\; 2\omega
\; A' \; . \nonumber
\end{eqnarray}

\noindent and
$$
 N \alpha + \beta M = \cos 2A \; , \qquad 2N\beta = \sin 2A\; , \qquad 2M \alpha = - \sin 2A \; ,
 $$
 $$
  \alpha \beta = \sin A \cos A ={1\over 2} \sin 2A \; ,
\qquad
 NM = - \sin A \cos A = - {1\over 2} \sin 2A \; ,
$$
$$
N^{2} = \cos^{2} A \; , \qquad \beta^{2} = \sin ^{2} A \; , \qquad
M^{2} = \sin^{2} A\;, \qquad \alpha^{2} = \cos^{2} A \; ,
$$

Therefore, eqs.  (\ref{8.11}) and (\ref{8.12}) take the  form
\begin{eqnarray}
2 \omega {dG_{2} \over dy} - \left [ \cos 2A {km \over y(1-y)} +
{1 \over 2}\; \sin 2A \;
 { - \omega^{2} y + m^{2} +
  \omega^{2}(1-y) - k^{2} \over y(1-y) } \; \right ] \; G_{2} +
\nonumber
\\[4mm]
+ \left [ + 2 \omega A' - \sin 2A \; {km \over y(1-y)} + \cos^{2}
A \; { - \omega^{2} y + m^{2} \over y(1 - y) } - \sin ^{2} A\;
{\omega^{2}(1-y) - k^{2} \over y(1-y)} \; \right ] \; G_{3} = 0 \;
, \nonumber
\end{eqnarray}
\begin{eqnarray}
2 \omega {dG_{3} \over dy} + \left [
  \cos 2A {km \over y(1-y)} + {1 \over 2} \; \sin 2A \; { -\omega^{2}y + m^{2} +
\omega^{2}(1-y) - k^{2} \over y(1-y)} \; \right ] \; G_{3} +
\nonumber
\\[4mm]
+ \left [ -2\omega A' - \sin 2A \; {km \over y(1-y)} - \sin^{2} A
\; {- \omega^{2} y + m^{2} \over y(1 - y) } + \cos ^{2} A \;
{\omega^{2}(1-y) - k^{2} \over y(1-y) } \; \right ] \; G_{2} = 0
\; , \nonumber
\end{eqnarray}

Supposing that the used linear transformation  does not depend on coordinate
 $y$, we get more simple expressions:
 \begin{eqnarray}
2 \omega {dG_{2} \over dy} - \left [ \cos 2A {km \over y(1-y)} +
{1 \over 2}\; \sin 2A \;
 { - \omega^{2} y + m^{2} +
  \omega^{2}(1-y) - k^{2} \over y(1-y) } \; \right ] \; G_{2} +
\nonumber
\\[4mm]
+ { - 2 km \sin 2A + (1 + \cos 2 A ) [ - \omega^{2} y + m^{2} ] -
(1 - \cos 2 A ) \; [ \omega^{2}(1-y) - k^{2} ] \over 2y(1-y)} \;
G_{3} = 0 \; , \label{8.14}
\end{eqnarray}

\begin{eqnarray}
2 \omega {dG_{3} \over dy} + \left [
  \cos 2A {km \over y(1-y)} + {1 \over 2} \; \sin 2A \; { -\omega^{2}y + m^{2} +
\omega^{2}(1-y) - k^{2} \over y(1-y)} \; \right ] \; G_{3} +
\nonumber
\\[4mm]
+ { - 2km \sin 2A - (1 - \cos 2 A ) [ - \omega^{2} y + m^{2}] + (1
+ \cos 2 A ) [ \omega^{2}(1-y) - k^{2} ] \over 2 y (1-y) } \;
G_{2} = 0 \; , \label{8.15}
\end{eqnarray}

Let
$$
\cos 2A = 0 \;, \qquad 2A = {\pi \over 2} \, \qquad \sin 2A = 1
$$

\noindent then eqs.  (\ref{8.14})--(\ref{8.15}) read
\begin{eqnarray}
\left (2 \omega {d \over dy} -
 { - \omega^{2} y +
  \omega^{2}(1-y) + m^{2} - k^{2} \over 2 y(1-y) } \right ) \; G_{2} +
{ - \omega^{2} + (m-k)^{2} \over 2y(1-y)} \; G_{3} = 0 \; ,
\nonumber
\end{eqnarray}

\begin{eqnarray}
\left (2 \omega {d \over dy} +
  { -\omega^{2}y +
\omega^{2}(1-y) + m^{2} - k^{2} \over 2 y(1-y)} \right ) \; G_{3}
+ { \omega^{2} - (m+k)^{2} \over 2y (1-y) } \; G_{2} = 0 \; .
\label{8.16}
\end{eqnarray}

It is remarkable that in the system produced the singularities are located  at
 the points $y = 0,1, \infty$) only.

From  (\ref{8.16}) and excluding $G_{3}$ one straightforwardly gets the equation  for  $G_{2}$:
\begin{eqnarray}
G_{3}=-2 \omega {2 y (1-y)\over -\omega^{2}+(m-k)^{2}}
{dG_{2}\over dy}+{- \omega^{2}y+\omega^{2}(1-y)+m^{2}-k^{2}\over
-\omega^{2}+(m-k)^{2}}G_{2},\nonumber
\\[4mm]
4 y (1 - y){d^{2}G_{2}\over dy^{2}}+4 (1 - 2 y){dG_{2}\over dy}+
 \left ( 2\omega +\omega^{2 }-{ m^{2}\over y(1-y)}+{
m^{2}-k^{2}\over 1-y} \right ) \; G_{2}=0 \label{8.18}
\end{eqnarray}

With the substitution$
G_{2} = y^{A} (1-y)^{B} G (y) $,
eq.  (\ref{8.18}) takes the form
\begin{eqnarray}
4 y (1-y) \; G '' + 4 \; \left [ A (1-y) \; - B y + A (1-y) - B y
\; + (1 -2y) \; \right ] \; \; G\; ' + \nonumber
\\
+ \left [\; 4 A (A-1) {1 \over y } + 4 B(B-1) {1 \over 1-y} - 4 A
(A-1)
 - 4 B(B-1) - 8 AB + \right.
\nonumber
\\
\left. + 4 ( -2A - 2B + {A \over y} + {B \over 1- y} ) +
 2\omega + \omega^{2 } -
   m^{2} ( {1 \over y} + {1 \over 1 - y} )
 + { m^{2}-k^{2}\over 1-y} \; \right ] \; G = 0
\label{8.20}
\end{eqnarray}

Requiring\begin{eqnarray}
4 A (A-1) + 4 A - m^{2} = 0 \; \; \Longrightarrow \; \; A= \pm \;
{1 \over 2} \mid m \mid \; , \nonumber
\\
4 B(B-1) + 4B - k^{2} = 0 \; \; \Longrightarrow \; \; B = \pm \;
{1 \over 2} \mid k \mid \; ;\;\;\; \label{8.21}
\end{eqnarray}

\noindent we arrive at
\begin{eqnarray}
 y (1-y) \; G '' + [ 2A+1 - 2 (A+B+1) y \; ] \; G\; ' -
\nonumber
\\[2mm]
- \left [\; (A +B) (A+B+1) - {\omega \over 2} ( {\omega \over 2} +
1 ) \; \right ] \; G = 0 \; ,
\nonumber
 \label{8.22}
\end{eqnarray}

\noindent what is of hypergeometric type
\begin{eqnarray}
\gamma = 2A +1 \; , \qquad \alpha + \beta = 2A + 2B+1 \;,
\nonumber
\\[3mm]
\alpha \beta = (A+B)(A+B+1) - {\omega \over 2} ({\omega \over 2} +
1) \; , \nonumber
\end{eqnarray}

\noindent that is \begin{eqnarray}
\alpha = A+B- {\omega \over 2} \;, \qquad \beta = A+B+1+ {\omega
\over 2} \;, \qquad \gamma = 2A +1 \; .
 \label{8.24}
\end{eqnarray}

The functions are finite on the  sphere $S_{3}$ only if
\begin{eqnarray}
A = + {1 \over 2} \mid m \mid \; , \qquad B = + {1 \over 2} \mid k
\mid \; , \qquad \alpha= - n = 0, -1, -2, ... \; ;
\nonumber
\label{8.25}
\end{eqnarray}

\noindent which leads to the frequency spectrum in the form :
\begin{eqnarray}
\omega = 2 ( n + A + B ) = 2n + \mid m \mid + \mid k \mid \; ;
\label{8.25}
\end{eqnarray}

\noindent the parameters $m$ and $k$  are allowed to be integer only :
$
m \; , \; k \in \{ 0, \pm 1, \pm 2, ...\; \}
$.
The function  $G_{2}(y)$ is
\begin{eqnarray}
G_{2} (y) = M_{2} \; y^{\mid m \mid/2} \; (1- y )^{\mid k \mid /2
}\; F ( -n , \; n +1 + \mid m \mid + \mid k \mid, \; \mid m \mid +
1 ; \; y ) \; . \label{8.27}
\end{eqnarray}

In eqs. (\ref{8.16}) we might exclude $G_{2}$:
$G_{3}$:
\begin{eqnarray}
G_{2} (y) =-2 \omega {2 y (1-y)\over \omega^{2}-(m+k)^{2}}
{dG_{3}\over dy}-{- \omega^{2}y+\omega^{2}(1-y)+m^{2}-k^{2}\over
\omega^{2}-(m+k)^{2}}G_{3},\nonumber
\\[4mm]
4 y (1 - y){d^{2}G_{3}\over dy^{2}}+4 (1 - 2 y){dG_{3}\over dy}+
 \left ( -2\omega +\omega^{2 }-{ m^{2}\over y(1-y)}+{
m^{2}-k^{2}\over 1-y} \right ) \; G_{3}=0 \label{8.29}
\end{eqnarray}

With the use of substitution
$G_{3} = y^{A} (1-y)^{B} F (y)$, the equation for $G_{3}$ reduces to
\begin{eqnarray}
4 y (1-y) \; F '' + 4 \; \left [ A (1-y) \; - B y + A (1-y) - B y
\; + (1 -2y) \; \right ] \; \; F\; ' + \nonumber
\\
+ \left [\; 4 A (A-1) {1 \over y } + 4 B(B-1) {1 \over 1-y} - 4 A
(A-1)
 - 4 B(B-1) - 8 AB + \right.
\nonumber
\\
\left. + 4 ( -2A - 2B + {A \over y} + {B \over 1- y} )
 -2\omega + \omega^{2 } -
   m^{2} ( {1 \over y} + {1 \over 1 - y} )
 + { m^{2}-k^{2}\over 1-y} \; \right ] \; F = 0
\label{8.31}
\end{eqnarray}

\noindent Requiring
\begin{eqnarray}
4 A (A-1) + 4 A - m^{2} = 0 \; \; \Longrightarrow \; \; A= + \; {1
\over 2} \mid m \mid \; , \nonumber
\\
4 B(B-1) + 4B - k^{2} = 0 \; \; \Longrightarrow \; \; B = + \; {1
\over 2} \mid k \mid \; ;\;\;\; \nonumber
\end{eqnarray}

\noindent we arrive at a hypergeometric type equation
\begin{eqnarray}
 y (1-y) \; F '' + [ 2A+1 - 2 (A+B+1) y \; ] \; F\; ' -
\nonumber
\\[2mm]
- \left [\; (A +B) (A+B+1) - {\omega \over 2} ( {\omega \over 2} -
1 ) \; \right ] \; F = 0 \; ,
\nonumber
\\
a = A+B+1- {\omega \over 2} \;, \qquad b = A+B+ {\omega \over 2}
\;, \qquad c = 2A +1 \; . \label{8.33}
\end{eqnarray}

\noindent
Further we get
\begin{eqnarray}
a = A+B+1- {\omega \over 2} = - N , \qquad N = 0, 1, 2, ... ,
\nonumber
\\
\omega = 2 (A + B +1 + N) = \mid m \mid + \mid k \mid + 2(1+N) \;
, \qquad \underline{ N +1 = n }\; , \nonumber
\\[3mm]
G_{3} = M_{3}
 \; y^{\mid m \mid/2} \; (1- y )^{\mid k \mid /2 }\; F ( -n +1 , \; n + \mid m \mid + \mid k \mid, \;
  \mid m \mid + 1 ; \; y ) \; ;
\label{8.34}
\end{eqnarray}

\noindent compare with  (\ref{8.27}).

It remains to find a relative factor in two functions
 $G_{2}$ and  $G_{3}$:
\begin{eqnarray}
G_{2} = M_{2} \; y^{\mid m \mid/2} \; (1- y )^{\mid k \mid /2 }\;
F ( -n , \; n +1 + \mid m \mid + \mid k \mid, \; \mid m \mid + 1 ;
\; y )  \; ,
\nonumber
\label{8.35}
\\
G_{3} = M_{3} \; y^{\mid m \mid/2} \; (1- y )^{\mid k \mid /2 }\;
F ( -n +1 , \; n + \mid m \mid + \mid k \mid, \; \mid m \mid + 1 ;
\; y ) \; , \label{8.36}
\end{eqnarray}

\noindent and the relationship  (see (\ref{8.18}))
\begin{eqnarray}
G_{3} \; [ (m-k)^{2} -\omega^{2} ] = -4 \omega \; y (1-y) \; {d
G_{2} \over dy} + [ m^{2}- k^{2} + \omega^{2} (1 -2y) ] \;G_{2} \;
\; .
\nonumber
\label{8.37}
\end{eqnarray}

\noindent must hold. Using the expressions for  $G_{2}$ Ё $G_{3}$ we get
\begin{eqnarray}
( m- k - \omega ) \; ( m- k + \omega) \; {M_{3} \over M_{2} }
 \; F_{3} (y) =
\nonumber
\\
=
 -4 \omega \; [ \;
{ \mid m \mid \over 2} \; (1-y) \; F_{2} (y) -
 {\mid k \mid \over 2 }\; y\; \; F_{2} (y)
 +
 \nonumber
 \\
 +
 y (1-y) \; {d \over dy} \; F_{2} (y) \; ] + [ m^{2}- k^{2} + \omega^{2} (1 -2y) ] \; F_{2} (y) \; .
\label{8.38}
\end{eqnarray}

It is sufficient to consider this equation   in the point $y=0$ only
that results in
\begin{eqnarray}
-( \omega + m- k ) \; ( \omega -m+ k ) \; \; {M_{3} \over M_{2} }
= (\omega - \mid m \mid - k ) \; (\omega - \mid m \mid + k ) \; ,
\nonumber
\end{eqnarray}

\noindent and therefore 
\begin{eqnarray}
M_{2} = M\; ( \omega + m - k) ( \omega - m + k) \; , \nonumber
\\
M_{3} = - M\; ( \omega - \mid m \mid - k) ( \omega - \mid m \mid +
k) \; . \label{8.39}
\end{eqnarray}

Depending on the sign of  $m$ it may be rewritten in a simpler form:
\begin{eqnarray}
m > 0\;, \qquad M_{2} = M (\omega - k + m ) \; , \qquad M_{3} = -
M (\omega - k -m ) \; ; \nonumber
\\[3mm]
m < 0\;, \qquad M_{2} = M (\omega + k - m ) \; , \qquad M_{3} = -
M (\omega + k + m ) \; ; \nonumber
\\
m=0\; , \qquad M_{2} = M\; , \qquad M_{3} = - M \; ; \label{8.40}
\end{eqnarray}

\noindent  $M$ stands for a numerical constant.

\section{ Maxwell solutions in elliptical  model }

\hspace{5mm} Let us consider  the problem of Maxwell solutions in
elliptical space $S\;'_{3}$.  This space  $S\;'_{3}$  is a space of  constant positive curvature also
and differs from the spherical model  in topological properties only:
  $S_{3}$  is 1-connected, $S\;'_{3}$ ia a 2-connected. The  question is on the role
  of these differences for electromagnetic field solutions.

To obtain explicit realizations for two models it is convenient to
use relations known in the theory of unitary and orthogonal groups.
To each point in $S_{3}$ there exists corresponding element in unitary group  $SU(2)$:
\begin{eqnarray}
B = \sigma ^{0}\; n_{0}\; - \;i \;\sigma ^{k} \; n_{k} \; ,\qquad
 \mbox{det} \; B= + 1  \;.
\nonumber
\end{eqnarray}

In turn, to each point in elliptic space $S\;'_{3}$  there exists
corresponding element in $SO(3)$ parameterized by Gibbs 3-vector
\cite{23}
\begin{eqnarray}
0(\vec{c}) =   \;I \;+\; 2 \; {\vec{c}^{\times } +
(\vec{c}^{\times })^{2}
 \over (1 + \vec{c}^{2})}\;   \; , \;\;
(\vec{c}^{\; \times })_{kl}  = - \; \epsilon _{klj} \; c_{j}   \;
. \label{9.1a}
\end{eqnarray}

\noindent note that  two infinite length  vectors  represent one
the same point in $S\; '_{3}$:
\begin{eqnarray}
\vec{c}^{\; +}_{\infty } = + \infty \; \vec{c}_{0}\;  , \qquad
\vec{c}^{\; -}_{\infty } = - \infty \;  \vec{c}_{0} \;, \qquad \;
\vec{c}^{\; 2}_{0} = 1 \; , \qquad
0(\vec{c}^{\; \pm \;
\infty  }) =  I \; + \; 2\; (\vec{c}^{\; \times }_{0})^{2}  \;.
\nonumber
\end{eqnarray}

Mapping $2\; \rightarrow \; 1$  from  $SU(2)$  to
$SO(3)$  is given by
\begin{eqnarray}
\{ + n_{a}\; ; \;\; - n_{a} \} \;\;   \rightarrow \;\;  \vec{c} =
{ \vec{n} \over n_{0} }\; .
\nonumber
\end{eqnarray}

Cylindric coordinates  $(\rho ,\phi ,z)$  in elliptic space can be  defined by the relations
\begin{eqnarray}
c_{1}  = { \tan\;  \rho  \over \cos  \rho } \; \cos  z \; ,
\; c_{2}  = { \tan\;  \rho \over  \cos  z }  \;  \sin  z \; ,
\; c_{3}  = \tan\;  z\; ,
\nonumber
\\
\tilde{G}\; , \qquad  \rho  \in  [ 0 , \pi /2 ] \; , \; \phi  \in  [-
\pi  ,\;  + \pi  ]\;  ,  \; z \in  [-\pi /2, +\pi /2 ] \; .
\label{9.2b}
\end{eqnarray}

\noindent Additionally, we must define such an identification rule on the boundary
of the region
$\tilde{G}$, which agrees with the identification rule  for vectors $\vec{c}^{+}_{\infty }$ and
$\vec{c}^{-}_{\infty }$.  To this end, it is convenient to divide the region
$\tilde{G}$  into three parts:
\begin{eqnarray}
\tilde{G}_{1} = \tilde{G} (\rho  \neq  0 , \pi /2) \;,\;\;
\tilde{G}_{2} = \tilde{G} (\rho = 0) \; , \;\; \tilde{G}_{3} =
\tilde{G} (\rho = \pi /2) \; .
\nonumber
\end{eqnarray}

For the region   $\tilde{G}_{1}$ identification is given by (for more detail see [...])

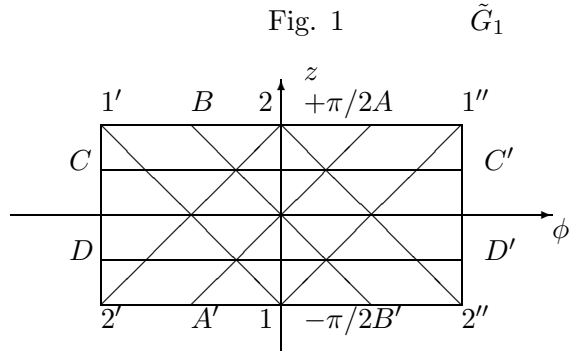
\begin{center}
Fig. 1 $\qquad \qquad \tilde{G}_{1}$
\end{center}

\vspace{2mm} \unitlength=0.6 mm
\begin{picture}(160,60)(-100,-30)
\special{em:linewidth 0.4pt} \linethickness{0.4pt}

\put(-60,0){\vector(+1,0){120}} \put(+60,-5){$\phi $}
\put(0,-30){\vector(0,+1){60}} \put(+5,+30){$z$}

\put(-40,-20){\line(+1,0){80}} \put(-40,-20){\line(0,+1){40}}
\put(+40,+20){\line(-1,0){80}} \put(+40,+20){\line(0,-1){40}}
\put(+40,+20){\line(-1,-1){40}} \put(-40,-20){\line(+1,+1){40}}
\put(-20,-20){\line(+1,+1){40}} \put(-40,+20){\line(+1,-1){40}}
\put(-20,+20){\line(+1,-1){40}} \put(0,+20){\line(+1,-1){40}}
\put(-40,-10){\line(+1,0){80}} \put(-40,+10){\line(+1,0){80}}
\put(-40,-25){$2'$} \put(-20,-25){$A'$}   \put(-5,-25){$1$}

\put(+5,-25){$-\pi /2$} \put(+20,-25){$B'$} \put(+40,-25){$2''$}

\put(-40,+23){$1'$}   \put(-20,+23){$B$}   \put(-5,+23){$2$}

\put(+5,+23){$+\pi /2$} \put(+20,+23){$A$} \put(+40,+23){$1''$}

\put(-47,+10){$C$}     \put(+45,+10){$C'$} \put(-47,-10){$D$}
\put(+45,-10){$D'$}

\end{picture}

\vspace{5mm}

\noindent  each pair $(A, A')\;  , \; (B, B')$  and so on represents one the same point in elliptical model $S'_{3}$;
also $(1, 1', 1'')$  and $(2, 2 ', 2'')$  correspond to one respective point in $S_{3}'$.

Now we should find what of above constructed Maxwell solutions  in case of spherical model
$S_{3}$ will be single-valued ones when considering elliptical model $S'_{3}$ (here we examine only points
parameterized by the region  $\tilde{G}_{1}$). Evidently, it is sufficient to examine the behavior of the vector
$f = e^{im\phi } \; e^{ikz}$. From the relations
\begin{eqnarray}
f(C)  = f(C') \; , \qquad  f(D) = f(D') \; , \;\;  ...
\nonumber
\end{eqnarray}

\noindent no  additional restrictions  arise besides that $M$ and $K$  to be integer.
Equations $f(1)  =
f(1' )  = f(1'' )$ give
\begin{eqnarray}
e^{-ik (\pi /2)}   = e^{-im\pi } \; e^{+ik (\pi /2)}   = e^{+im\pi
}\; e^{+ik(\pi /2)} \; ,
\nonumber
\end{eqnarray}

\noindent  whence it follows that
\begin{eqnarray}
e^{i2m\pi } = 1\; ,\qquad   e^{i(k-m) \pi } = 1\; ,\qquad
e^{i(k+m) \pi } = 1 \; ;
\nonumber
\end{eqnarray}

\noindent and therefore,  $( k - m )$  and $( k + m )$ must be even. The same results follows  from
consideration of points  $(2 ,2', 2'')$.

Therefore, Maxwell solutions constructed above in  spherical space  will be single-valued  solutions
 in elliptical space (in region  $\tilde{G}_{1}$ ) only if  $m$ and $k$ are both even, or both odd;
 correspondingly,
$\omega$  parameter  $N$  in the expression for the frequency spectrum (see.  (\ref{8.25})) given by
\begin{eqnarray}
\omega = 2n + \mid m \mid + \mid k \mid = N
\label{9.8}
\end{eqnarray}

\noindent takes on  even values: $ N = 0, 2,\; 4,\; 6,\;
\ldots $

Now let us consider the behavior of the above mentioned Maxwell solutions in the remaining regions
$\tilde{G}_{2}$  and  $\tilde{G}_{3}$.

First, let us specify the case of   $\tilde{G}_{2}$ and consider the vicinity of the point  $P$ (see Fig. 2):

$$
\rho  = ( \pi /2\; \tan\;\alpha  + z\; \tan\; \alpha
) \;\;\; \Longrightarrow \;\;\; \{ z = -\pi /2 + \delta  , \rho  = \delta \;
\tan\; \alpha  \} \; ;
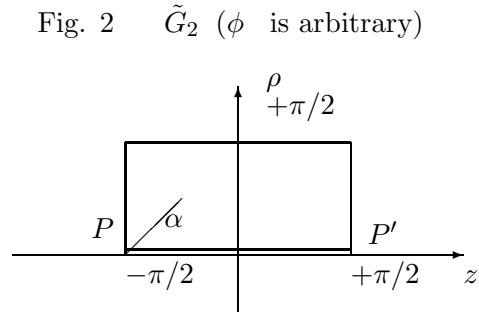
$$
\vspace{5mm}

\begin{center}
Fig. 2 $\;\;\;\; \;  \tilde{G}_{2} \;\;(\phi \;\;$ is arbitrary)
\end{center}

\vspace{2mm} \unitlength=0.75mm
\begin{picture}(160,60)(-100,-30)
\special{em:linewidth 0.4pt} \linethickness{0.4pt}

\put(-26,+3){$P$}       \put(+23,+2){$P'$}

\put(-40,0){\vector(+1,0){80}} \put(+40,-5){$z$}

\put(0,-10){\vector(0,+1){40}} \put(+5,+30){$\rho $}

\put(-20,0){\line(0,+1){20}}    \put(+20,0){\line(0,+1){20}}

\put(-20,+20){\line(+1,0){40}}  \put(-20,+1){\line(+1,0){40}}

\put(-20,0){\line(+1,+1){10}} \put(-13,+5){$\alpha $}

\put(-20,-5){$-\pi /2$} \put(+20,-5){$+\pi /2$} \put(+5,+25){$+\pi
/2$}

\end{picture}
\vspace{-15mm}

\noindent and
\begin{eqnarray}
c_{1} = { \tan\;  (\delta  \; \tan\;  \alpha ) \over
\sin ( \delta  \; \tan\; \alpha )} \;{ \cos  \phi \over \sin
\delta }\;  , \;\; c_{2} = { \tan\;  (\delta  \; \tan\;
\alpha ) \over \sin ( \delta \;\tan\; \alpha )} \; {\sin \phi
\over \sin \delta }\; ,\;\; c_{3} = - {\cos  \delta  \over \sin
\delta } \; ;
\nonumber
\end{eqnarray}

\noindent so in the limit  $\delta \rightarrow 0\; ( \alpha  \neq
0)$ we get
\begin{eqnarray}
\tilde{G}_{2} , \qquad P \; , \qquad \vec{c} = + \; \infty \;  ( 0
\; ,\;  0\;  , \; - 1 ) \; .
\label{9.9a}
\end{eqnarray}

\noindent This means that coordinate  $\phi $  is "mute"   \hspace{2mm}
at the point $( \rho  = 0 ,\; z = - \pi /2, \;\phi  )$.

For another point  $P '  = ( \rho  = 0 ,\; z = + \pi /2, \;\phi  )$ we get similar result:
\begin{eqnarray}
\tilde{G}_{2}\; , \qquad  P' \;,  \qquad \vec{c} = - \; \infty \;
( 0 \;, \;0 \;,\;- 1 ) \; .
\label{9.9b}
\end{eqnarray}

\noindent Thus,  $P$   and  $P'$  represent one the same point in elliptical  space.
Compare  $\Phi_{\omega  mk}$  at these two points $P$  and $P'$:
\begin{eqnarray}
\Phi _{\omega m k} \sim e^{im \phi } \;e^{ik z} \; (\sin r )^{\mid
m \mid } \; (\cos r )^{\mid k \mid } F(A,B,C;  \sin^{2} r ) \; ;
\nonumber
\end{eqnarray}

\noindent  and
\begin{eqnarray}
\Phi _{\omega m k}(P) \sim  \left \{ \begin{array}{l}
0 ,  \;\; \mbox{if} \;\; m \neq  0 ; \\
e^{+ik\pi /2}\; F(A, B, C; 0), \;\;  \mbox{if} \;\; m = 0 ;
\end{array} \right.
\nonumber
\\
\Phi _{\omega m k}(P')  \sim  \left \{ \begin{array}{l}
 0   , \;\; \mbox{if} \;\; m \neq  0   ; \\
e^{-ik\pi /2}\; F(A, B, C; 0),\;\; \mbox{if} \;\; m = 0 \; ;
\end{array} \right.
\label{9.10}
\end{eqnarray}

\noindent because $k$ is  even (at $m=0$) we have the identity
 $\Phi _{\omega m k}(P)  = \Phi
_{ \omega m k}(P')$. In remaining part of the region
$\tilde{G}_{2}$,  points of elliptical model  are parameterized  according to
\begin{eqnarray}
( 0 ; \phi  ; z \neq  0, \pi \neq 2 ) \; \Longrightarrow \qquad
\nonumber
\\
 \vec{c} = ( 0\; ,\; 0\; , \tan \;z ) \;,\;
\tilde{G}_{2}\;\; (\phi  - \mbox{"mute" \hspace{2mm} variable}
)  \; .\label{6.3.11a}
\end{eqnarray}

\noindent and the function  $\Phi _{\omega m k }$
\begin{eqnarray}
G_{2}\; , \qquad \phi _{\omega m k} \sim  \left \{
\begin{array}{l}
0 , \;\; if \;\; m \neq  0 ; \\
e^{ikz}\; F (A, B, C; 0) \; , \;\; \mbox{if} \;\; m = 0\; ,
\end{array} \right.
\label{13.3.11b}
\end{eqnarray}

\noindent is single-valued and continuous in that part of   $\tilde{G}_{2}$ of the elliptic space.

Finally, let us  consider the region   $\tilde{G}_{3}$:

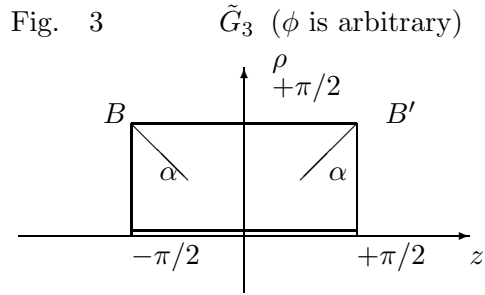
\begin{center}
Fig. $\;\; 3 \qquad \qquad \tilde{G}_{3} \;\; (\phi$  is arbitrary)
\end{center}

\unitlength=0.75 mm
\begin{picture}(160,60)(-100,-30)
\special{em:linewidth 0.4pt} \linethickness{0.4pt}

\put(-25,+20){$B$}       \put(+25,+20){$B'$}
\put(-40,0){\vector(+1,0){80}} \put(+40,-5){$z$}
\put(0,-10){\vector(0,+1){40}} \put(+5,+30){$\rho $}

\put(-20,0){\line(0,+1){20}}    \put(+20,0){\line(0,+1){20}}
\put(-20,+20){\line(+1,0){40}}  \put(-20,+1){\line(+1,0){40}}

\put(-20,+20){\line(+1,-1){10}} \put(-15,+10){$\alpha $}

\put(+20,+20){\line(-1,-1){10}} \put(+15,+10){$\alpha $}

\put(-20,-5){$-\pi /2$} \put(+20,-5){$+\pi /2$} \put(+5,+25){$+\pi
/2$}

\end{picture}
\vspace{-10mm}

\noindent In the vicinity of  $B$ we have
\begin{eqnarray}
\rho = [\; - z\; \tan\; \alpha + ({\pi \over 2} - {\pi \over
2}\;\tan\; \alpha )\; ] \;\rightarrow \; \{ \; z = (- {\pi
\over 2} + \delta  )\; ,\; \rho  = ({\pi \over 2}
 - \delta \; \tan\; \alpha  )\; \}  \; ;
\nonumber
\end{eqnarray}

\noindent that results in
\begin{eqnarray}
c_{1} = { \cos  (\delta \; \tan\;  \alpha  ) \over \sin
(\delta \; \tan\; \alpha ) } \;{ \cos \phi \over \sin \delta
}\; , \;\; c_{2} = { \cos (\delta \; \tan\; \alpha ) \over
\sin (\delta \; \tan\; \alpha )} \; { \sin  \phi  \over \sin
\delta }\; ,\;\; c_{3} = - { \cos  \delta  \over \sin  \delta }
\;;
\nonumber
\end{eqnarray}

\noindent from whence in the limit  $ \delta \; \rightarrow \; 0 $ (as
$\alpha  \neq  \pi /2$ ) it follows
\begin{eqnarray}
B\; : \qquad \vec{c} = \infty  \; {1 \over \tan\; \alpha } \;
(\cos \phi , \; \sin \phi , \; 0 ) = \infty  \; (\cos \phi , \;
\sin \phi , \; 0 ) \; .
\nonumber
\end{eqnarray}

\noindent  Analogously, in the vicinity of  $B'$  we get
\begin{eqnarray}
B' \;  : \qquad \vec{c} = \infty  \; (\cos \phi , \; \sin \phi ,
\; 0 ) \; .
\nonumber
\end{eqnarray}

\noindent Comparing function at  points  $B$  and $B'$:
\begin{eqnarray}
\Phi _{\omega m k }(B)  \sim   \left \{ \begin{array}{l}
 0 \; ,  \;\;  \mbox{if}\;\; k \neq  0  ; \\
e^{+i m \pi /2}\; F (A, B, C; 1) \;, \;\; \mbox{if}\;\; k = 0 \;
;
\end{array} \right.
\nonumber
\\
\Phi _{\omega m k }(B') \sim   \left \{ \begin{array}{l}
0  \;, \;\;   \mbox{if} \;\; k \neq  0\;   ; \\
e^{-im \pi /2} \; F (A, B, C; 1) \;,\;\; \mbox{if} \;\; k = 0\;
;
\end{array} \right.
\label{9.12}
\end{eqnarray}

\noindent because  $m$ is even,  the equality
 $ \Phi _{\omega m k }(B)  = \Phi _{\omega mk}(B')$ holds.
In remaining part of   $\tilde{G}_{3}$:
\begin{eqnarray}
\tilde{G}_{3} \;:\qquad  (\pi /2, \; \phi , \; z \neq  - \pi /2, +
\pi /2) \; \Longrightarrow \;
\nonumber
\\
\vec{c} = { \infty  \over \cos  z } \; (\cos  \phi , \; \sin  \phi
, \; 0) = \infty \;(\cos  \phi , \; \sin  \phi , \; 0)\; .
\label{9.13a}
\end{eqnarray}

\noindent  $z$ is "mute"  \hspace{1mm} coordinate, therefore the solutions
\begin{eqnarray}
\tilde{G_{3}} = \left \{ \begin{array}{l}
 0 \;   ,\;\;  \mbox{if} \;\; k \neq  0\;  , \\
e^{imz}\; F (A, B, C; 1) \;  , \;\; \mbox{if} \;\; k = 0\;\;  ,
\end{array} \right.
\label{9.13b}
\end{eqnarray}

\noindent represent single-valued and continuous  functions
in that part of elliptic space.

Thus, all functions $\Phi _{\omega mk }(\rho ,\phi ,z)$ at $\omega
=  N = 0,\;2,\; 4,\; \ldots $  are single-valued  and continuous
in elliptic space, and they represent physical solutions for
Maxwell equation in this space, whereas all remaining functions
$\Phi _{\omega mk}(\rho ,\phi ,z) \;, \; N = 1,\; 3,\; \ldots $
should be rejected as non-single-valued and discontinuous in $S\;
'_{3}$ space.

\section{ Cylindric coordinate and tetrad in Lobachevsky space $H_{3}$,
}

Let us consider Maxwell equations in  cylindric  coordinate \cite{22} of
hyperbolic Lobachevsky space $H_{3}$:
\begin{eqnarray}
n_{1} = \sinh r \; \cos \phi \; , \;\; n_{2} = \sinh r \; \sin \phi \;
, \;\;
 n_{3} = \cosh r \; \sinh z \; , \;\; n_{4} =
\cosh r \; \cosh z \; ; \nonumber
\\
 d S^{2} = dt^{2} - d r^{2} -
\sinh^{2} r \; d\phi^{2} - \cosh^{2}r \; dz^{2} \; , \qquad x^{\alpha}
= (t, r , \phi, z)\; , \nonumber
\\
 e_{(a)}^{\beta}(y) = \left |
\begin{array}{llll}
1 & 0 & 0 & 0 \\
0 & 1 & 0 & 0 \\
0 & 0 & \sinh^{-1} r & 0 \\
0 & 0 & 0 & \cosh^{-1}r
\end{array} \right | \; , \;
 e_{(a) \beta}(y) = \left |
\begin{array}{llll}
1 & 0 & 0 & 0 \\
0 & -1 & 0 & 0 \\
0 & 0 & - \sinh r & 0 \\
0 & 0 & 0 & - \cosh r
\end{array} \right | \; ;
\label{10.1}
\end{eqnarray}

\noindent where  $(r,\; \phi,\; z )$ run within
\begin{eqnarray}
r \in [ 0 , + \infty )\; , \qquad \phi \in [ 0 , 2\pi ]\;, \qquad
z \in (- \infty , + \infty ) \; . \nonumber
\end{eqnarray}

Christoffel symbols are
\begin{eqnarray}
\Gamma^{0}_{\beta \sigma} = 0 \; , \qquad \Gamma^{i}_{00} = 0 \; ,
\qquad \Gamma^{i}_{0j} = 0 \; , \qquad
\Gamma^{z}_{\;\; jk} = \left | \begin{array}{ccc}
0 & 0 & {\sinh r \over \cosh r} \\
0 & 0 & 0 \\
{\sinh r \over \cosh r} & 0 & 0
\end{array} \right |,
\nonumber
\\
\Gamma^{r}_{\;\; jk} = \left | \begin{array}{ccc}
0 & 0 & 0 \\
0 & - \sinh r \cosh r & 0 \\
0 & 0 & -\sinh r \cosh r
\end{array} \right | ,
\Gamma^{\phi}_{\;\; jk} = \left | \begin{array}{ccc}
0 & {\cosh r \over \sinh r} & 0 \\
{\cosh r \over \sinh r} & 0 & 0 \\
0 & 0 & 0
\end{array} \right |   \; .
\nonumber
\end{eqnarray}

\noindent Derivatives of tetrad vectors read
\begin{eqnarray}
e_{(0) \beta ; \alpha }= {\partial e_{(0)\beta} \over \partial
x^{\alpha}} - \Gamma^{\sigma} _{\alpha \beta} e_{(0)\sigma}  =  0 \; ,
\nonumber
\end{eqnarray}
\begin{eqnarray}
e_{(1) \beta ; \alpha }= {\partial e_{(1)\beta} \over \partial
x^{\alpha}} - \Gamma^{r} _{\alpha \beta} \; e_{(1)r} =  \left | \begin{array}{cccc}
0 & 0 & 0 & 0 \\
0 & 0 & 0 \\
0 & 0 & - \sinh r \cosh r & 0 \\
0 & 0 & 0 & -\sinh r \cosh r
\end{array} \right | \; ,
\nonumber
\end{eqnarray}
\begin{eqnarray}
e_{(2) \beta ; \alpha }= {\partial e_{(2)\beta} \over \partial
x^{\alpha}} - \Gamma^{\phi} _{\alpha \beta} \; e_{(2)\phi} =  \left | \begin{array}{cccc}
0 & 0 & 0 & 0 \\
0 & 0 & \cosh r & 0 \\
0 & 0 & 0 & 0 \\
0 & 0 & 0 & 0
\end{array} \right | ,
\nonumber
\end{eqnarray}
\begin{eqnarray}
e_{(3) \beta ; \alpha }= {\partial e_{(3)\beta} \over \partial
x^{\alpha}} - \Gamma^{z} _{\alpha \beta} \; e_{(3) z } = \left | \begin{array}{cccc}
0 & 0 & 0 & 0 \\
0 & 0 & 0 & \sinh r \\
0 & 0 & 0 & 0 \\
0 & 0 & 0 & 0
\end{array} \right | .
\label{10.5}
\end{eqnarray}

Ricci rotation coefficients are
\begin{eqnarray}
\gamma_{01 1}= \gamma_{02 1}= \gamma_{03 1} = 0 \; , \;
\gamma_{012}= \gamma_{022}= \gamma_{032} = 0 \; , \; \gamma_{01
3}= \gamma_{02 3}= \gamma_{03 3} = 0 \; ,
\nonumber
\end{eqnarray}

\noindent and
\begin{eqnarray}
\gamma_{23 1} =
0 \; , \; \; \gamma_{31 1} =   0 \; , \; \; \gamma_{12 1} =0 \; , \nonumber
\\
\gamma_{23 2}  = 0 \; , \;\; \gamma_{31 2} =  0 \; , \;\; \gamma_{12 2}
=  { \cosh r \over \sinh r} \; ,
\nonumber
\\
\gamma_{23 3} =  0 \; , \;\; \gamma_{31 3}  =
   -{\sinh r \over \cosh r }\; ,
\;\; \gamma_{12 3} =  0 \; . \nonumber
\end{eqnarray}

With  relations
\begin{eqnarray}
e_{(0)}^{\rho} \partial_{\rho} = \partial_{(0)} = \partial_{t} \;
, \qquad e_{(1)}^{\rho} \partial_{\rho} = \partial_{(1)} =
\partial_{r}\; ,
\nonumber
\\
e_{(2)}^{\rho} \partial_{\rho} = \partial_{(2)} = {1 \over \sinh r}
\partial_{\phi} \; , \qquad e_{(3)}^{\rho} \partial_{\rho} =
\partial_{(3)} = {1 \over \cosh r} \partial_{z} \; ,
\nonumber
\\
 {\bf v}_{0} =( \gamma_{01 0}, \gamma_{02 0} , \gamma_{03 0} ) \equiv 0 \; , \qquad
 {\bf v}_{1} = ( \gamma_{01 1}, \gamma_{021 } , \gamma_{03 1} ) \equiv 0 \; ,
\nonumber
\\
 {\bf v}_{2} = ( \gamma_{010}, \gamma_{02 2} , \gamma_{03 2} ) \equiv 0 \; ,
\qquad
 {\bf v}_{3} = ( \gamma_{013}, \gamma_{02 3} , \gamma_{03 3} ) \equiv 0 \; ,
\nonumber
\\
 {\bf p}_{0} = ( \gamma_{23 0}, \gamma_{31 0} , \gamma_{12 0} ) = 0 \; , \qquad
 {\bf p}_{1} = ( \gamma_{23 1}, \gamma_{31 1} , \gamma_{12 1} ) = 0 \; ,
 \nonumber
\\
 {\bf p}_{2} = ( \gamma_{23 2}, \gamma_{31 2} , \gamma_{12 2} ) = (0 , 0, {\cosh r \over \sinh r}) \; ,
\nonumber
\\
 {\bf p}_{3} = ( \gamma_{23 3}, \gamma_{31 3} , \gamma_{12 3} ) = (0, -{\sinh r \over \cosh r} , 0 ) \;
,
\nonumber
\end{eqnarray}

\noindent Maxwell equation (\ref{3.5}) reads
\begin{eqnarray}
\left ( - i \partial_{t} + \alpha^{1} \; \partial_{r} +
\alpha^{2}  {1 \over \sinh r} \partial_{\phi} +
 \alpha^{3} {1 \over \cosh r}\partial_{z}
 + \alpha^{2}  S_{3}  {\cosh r \over \sinh r} -\alpha^{3}  S_{2}  {\sinh r \over \cosh r}  \right )
 \left | \begin{array}{c}
0 \\ {\bf E} + i c{\bf B}
\end{array} \right | = 0  .
\label{10.11}
\end{eqnarray}

\section{ Separation of variables in space $H_{3}$ }

Maxwell matrix operator in (\ref{10.11}) commutes with three operators:
 $i\partial_{t},\; i\partial_{\phi},\;i\partial_{z}$; therefore solutions can be constructed on the base of
 the following  substitution
\begin{eqnarray}
\Psi = \left | \begin{array}{c} 0 \\ {\bf E} + i c{\bf B}
\end{array} \right | =
 e^{-i \omega t} \; e^{im\phi} \; e^{ikz} \;
 \left |
\begin{array}{c} 0 \\ f_{1}(r) \\ f_{2}(r) \\ f_{3}(r)
\end{array} \right | \; .
\label{11.2}
\end{eqnarray}

\noindent and  eq.  (\ref{10.11}) takes the form
\begin{eqnarray}
\left (\; - \omega + \alpha^{1} \; {d \over dr} + {im \over \sinh r}
\; \alpha^{2} + { ik \over \cosh r} \; \alpha^{3} + {\cosh r \over \sinh r} \; \alpha^{2} S_{3} - {\sinh r \over \cosh r}\; \alpha^{3} S_{2} \;
\right ) \left |
\begin{array}{c} 0 \\ f_{1}(r) \\ f_{2}(r) \\ f_{3}(r)
\end{array} \right |
 = 0 \; .
\label{11.3}
\end{eqnarray}

\noindent After simple calculations we get the radial system in the form
\begin{eqnarray}
({d \over dr } + {\cosh r \over \sinh r} + {\sinh r \over \cosh r }) f_{1}
+
 {im \over \sinh r}\; f_{2} + {ik \over \cosh r }\; f_{3} = 0 \; ,
\nonumber
\\
- \omega f_{1} - {ik \over \cosh r}\; f_{2} + {im \over \sinh r}\;
f_{3} = 0 \; , \nonumber
\\
-\omega f_{2} - ({d \over dr} + {\sinh r \over \cosh r} )\; f_{3} +
{ik \over \cosh r} f_{1} = 0 \; , \nonumber
\\
-\omega f_{3} + ({d \over dr} + {\cosh r \over \sinh r} )\; f_{2} -
{im \over \sinh r} f_{1} = 0 \; . \label{11.5}
\end{eqnarray}

\section{ Solutions at $m=0$, in space  $H_{3}$}

First, let  $m=0$ -- then
\begin{eqnarray}
({d \over dr } + {\cosh r \over \sinh r} + {\sinh r \over \cosh r }) f_{1}
 + {ik \over \cosh r }\; f_{3} = 0 \; ,
\nonumber
\\
 f_{1} = {-ik \over \omega \; \cosh r}\; f_{2} \; ,
\nonumber
\\
- \omega f_{2} - ({d \over dr} + {\sinh r \over \cosh r} )\; f_{3} +
{ik \over \cosh r} f_{1} = 0 \; , \nonumber
\\
f_{3} = {1 \over \omega}\; ({d \over dr} + {\cosh r \over \sinh r} )\;
f_{2} \; . \label{11.7}
\end{eqnarray}

\noindent With the help of 2nd and  4th equation from the
1st and 3rd it follows the identity $0=0$ and equation for $f_{2}$:
\begin{eqnarray}
f_{2}(r) = {1 \over \sinh r}\; E (r) \; , \qquad
{d^{2} E \over dr^{2}} - {1 \over \sinh r \cosh r}\; {dE \over dr} + (
\omega^{2} - {k^{2} \over \cosh^{2} r} ) E = 0\; ; \label{11.10a}
\end{eqnarray}

\noindent besides
\begin{eqnarray}
 f_{1} (r) = {-ik \over \omega } \; {1 \over \cosh r \sinh r }\; E(r) \; , \qquad
f_{3} = {1 \over \omega}\; {1 \over \sinh r} \; {d \over dr }E(r) \;.
\nonumber
\end{eqnarray}

\noindent Eq. (\ref{11.10a}) can be resolved very easily when $k^{2}=
\omega^{2}$:
\begin{eqnarray}
 {\sinh\; r \over \cosh\; r} {d \over dr} { \cosh\; r \over \sinh \; r} {d \over dr} \; E
    + k^{2} (1 - {1 \over \cosh^{2} r} ) \; E = 0 \; ;
\nonumber
\end{eqnarray}

\noindent from whence it follows
\begin{eqnarray}
( { \cosh\; r \over \sinh \; r} {d \over dr} )\;
 ( { \cosh \; r \over \sinh\; r} {d \over dr} ) \; E = -k^{2}\; E \; .
\label{12.2}
\end{eqnarray}

\noindent With the help of a new variable
\begin{eqnarray}
{ \cosh\; r \over \sinh\; r } { d \over dr } = {d \over dx
} \; , \qquad \Longrightarrow \qquad {d r \over dx } = {
\cosh\; r \over \sinh\; r } \; , \nonumber
\\
 dx = d \log \cosh\; r \;, \qquad x =
  \log ( C \; \cosh\; r ) \; , \qquad C = \mbox{const} \; .
\nonumber
\end{eqnarray}

\noindent  we arrive at
\begin{eqnarray}
{d^{2} \over dx^{2} } E = - k^{2} \; E \; , \qquad
 E = e^{i k x } = \mbox{const} \; ( \cosh\; r )^{i k}\; \; , \qquad k = \pm \; \omega \; ;
\nonumber
\label{12.3c}
\end{eqnarray}

\noindent Two constructed solutions are conjugated:
\begin{eqnarray}
( \cosh\; r )^{ + i k } = ( e ^{\log \cosh\; r} )^{+ i k }
= \cos (\; k \log \cosh\; r \; ) + i\; \sin (\; k \log
\cosh\; r \; ) \; , \nonumber
\\
 ( \cosh\; r )^{ -ik } = [ e ^{\log \cosh\; r} ]^{- i k } =
 \cos (\; k \log \cosh\; r \; ] - i\; \sin (\; k \log \cosh\; r \;) \; ,
\nonumber
\label{12.4}
\end{eqnarray}

\noindent therefore one can separate two independent real ones:
\begin{eqnarray}
E_{+} (r) = \cos \; [\; k_{0} \log \cosh\; r \; ] \; , \qquad
E_{-} (r) = \sin \; [\; k_{0} \log \cosh\; r \; ] \; .
\label{12.5a}
\end{eqnarray}

\noindent In the limit of vanishing curvature  they reduces to the known ones :
\begin{eqnarray}
r \rightarrow 0\;, \qquad E_{+} (r) = \cos \; (\; k_{0} \log
\cosh\; r \; ) \;\; \longrightarrow \;\; +1 \; ; \nonumber
\\
r \rightarrow 0\;, \qquad E_{-} (r) = \sin \; (\; k_{0} \log
\cosh\; r \; ) \;\; \longrightarrow \;\; \mbox{const}\; r^{2}
\; ; \label{12.5b}
\end{eqnarray}

\noindent these satisfies to the equation in the flat space-time
\begin{eqnarray}
 {d \over d r} { 1 \over r }\; {d \over d r}\; E(r) = 0 \;, \qquad \Longrightarrow \qquad
 E(r) \sim 1 , \; r^{2} \; .
\nonumber
\label{12.5c}
\end{eqnarray}

\noindent In contrast to the flat space,
 in Lobachevsky model the
  waves  (\ref{12.5a}) are  both oscillating at infinity.

  Let us show that (\ref{11.10a}) has another simple exact solution
  in the form
\begin{eqnarray}
E = \sinh^{2} r \; \cosh^{B}r \;  . \label{12.6b}
\end{eqnarray}

Indeed, substitution of (\ref{12.6b}) into (\ref{11.10a}) leads to
\begin{eqnarray}
2 \; \cosh^{4}r + 2(B+1) \; \sinh^{2} r\; \cosh^{2}r
+3B\; \sinh^{2}r \; \cosh^{2}r + B(B-1) \; \sinh^{4}r
- \nonumber
\\
- 2 \; \cosh^{2} r - B \; \sinh^{2} r \; - k^{2} \;
\sinh^{2} r + \omega^{2} \; \sinh^{2} r \; \cosh^{2}r
=0 \;. \nonumber
\end{eqnarray}

\noindent With notation $\cosh^{2} r = x \;  $ it reads
\begin{eqnarray}
x^{2} \; ( 4 + 4B + \omega^{2} + B^{2} ) +
  x \; ( -4 - 4B - \omega^{2} - B^{2} ) +
x^{0} \; ( B^{2} + k^{2} ) = 0 \; . \nonumber
\end{eqnarray}

\noindent The latter is satisfied if
\begin{eqnarray}
B^{2} = - k^{2} \; , \qquad (B +2)^{2} + \omega^{2} = 0 \; ,
\label{12.7a}
\end{eqnarray}

\noindent that is
\begin{eqnarray}
B = -2 + i\; \omega , \; -2 - i\; \omega \; , \qquad
 k = \pm\; i\; B=
 \left \{ \begin{array}{l}
\mp \; (2i + \omega ) \; ,
\\
\mp \; ( 2i - \omega ) \; ,
\end{array} \right.
\label{12.7b}
\end{eqnarray}

\noindent Corresponding solutions look as
\begin{eqnarray}
E(t,r,z) = E_{0} \; \sinh^{2} r \; \; \cosh^{B}r \; e^{-i
(\omega t - k z)} \;  .
\label{12.7c}
\end{eqnarray}

\noindent their  real and  imaginary parts are given by

\vspace{3mm} $B=-2 + i \omega,\;k=iB=-2i-\omega,$
\begin{eqnarray}
E(t,r.z) =E_{0} \; \sinh^{2} r \; \; \cosh^{-2 + i\;
\omega}r \; e^{i (-2zi-\omega z-\omega t )}= E_{0} \;
\tanh^{2} r \; \; \cosh^{ i \omega}r \;e^{2z} e^{i
(-\omega z-\omega t )}= \nonumber
\\
= E_{0} \; \tanh^{2} r \;e^{2z} \;[\cos (\omega \log \cosh r)+i\sin(\omega \log \cosh r)][\cos(-\omega z -\omega t)+i
\sin(-\omega z -\omega t)]= \nonumber
\\
=E_{0} \; \tanh^{2} r \;e^{2z}\cos (\omega \log \cosh r-\omega
z-\omega t)+i E_{0} \; \tanh^{2} r \; e^{2z}\sin(\omega \log
\cosh r-\omega z-\omega t) \; \nonumber
\end{eqnarray}

$ B=-2 + i \omega,\;k=-iB=2i+\omega, $
\begin{eqnarray}
E(t,r.z) =E_{0} \; \sinh^{2} r \; \; \cosh^{-2 + i\;
\omega}r \; e^{i (2zi+\omega z-\omega t )}= E_{0} \; \tanh^{2}
r \; \; \cosh^{ i \omega}r \;e^{-2z} e^{i (\omega z-\omega t
)}= \nonumber
\\
= E_{0} \; \tanh^{2} r \;e^{-2z} \;[\cos (\omega \log \cosh r)+i\sin(\omega \log \cosh r)][\cos(\omega z -\omega t)+i \sin(\omega
z -\omega t)]= \nonumber
\\
=E_{0} \; \tanh^{2} r \;e^{-2z}\cos (\omega \log \cosh r+\omega
z-\omega t)+i E_{0} \; \tanh^{2} r \; e^{-2z}\sin(\omega \log
\cosh r+\omega z-\omega t) \; \nonumber
\end{eqnarray}

$ B=-2 - i \omega,\;k=iB=-2i+\omega,$
\begin{eqnarray}
E(t,r.z) =E_{0} \; \sinh^{2} r \; \; \cosh^{-2 - i\;
\omega}r \; e^{i (-2zi+\omega z-\omega t )}= E_{0} \;
\tanh^{2} r \; \; \cosh^{ - i \omega}r \;e^{2z} e^{i
(\omega z-\omega t )}= \nonumber
\\
= E_{0} \; \tanh^{2} r \;e^{2z} \;[\cos (-\omega \log \cosh r)+i\sin(-\omega \log \cosh r)][\cos(\omega z -\omega t)+i
\sin(\omega z -\omega t)]= \nonumber
\\
=E_{0} \; \tanh^{2} r \;e^{2z}\cos (-\omega \log \cosh r+\omega
z-\omega t)+i E_{0} \; \tanh^{2} r \; e^{2z}\sin(-\omega \log
\cosh r+\omega z-\omega t) \;  \nonumber
\end{eqnarray}

$B=-2 - i \omega,\;k=-iB=2i-\omega,$]
\begin{eqnarray}
E(t,r.z) =E_{0} \; \sinh^{2} r \; \; \cosh^{-2 - i\;
\omega}r \; e^{i (2zi-\omega z-\omega t )}= E_{0} \; \tanh^{2}
r \; \; \cosh^{ - i \omega}r \;e^{-2z} e^{i (-\omega z-\omega
t )}= \nonumber
\\
= E_{0} \; \tanh^{2} r \;e^{-2z} \;[\cos (-\omega \log \cosh r)+i\sin(-\omega \log \cosh r)][\cos(-\omega z -\omega t)+i
\sin(-\omega z -\omega t)]= \nonumber
\\
=E_{0} \; \tanh^{2} r \;e^{-2z}\cos (-\omega \log \cosh r-\omega
z-\omega t)+i E_{0} \; \tanh^{2} r \; e^{-2z}\sin(-\omega \log
\cosh r-\omega z-\omega t) \; .
\nonumber
\end{eqnarray}

Physical sense of these  waves is not clear. More insight can be reached when
constructing all possible solutions of eq.  (\ref{11.10a}) through the substitution
\begin{eqnarray}
E = \sinh^{2} r \; \cosh^{B}r \; F (r)\; .
\nonumber
\end{eqnarray}

\noindent In that way, from (\ref{11.10a}) we arrive at an equation of hypergeometric type
(if  $ x = \sinh\;r, \;\; k^{2} + B^{2} = 0 $)
\begin{eqnarray}
4x (1-x) {d^{2} \over dx^{2}} F + 4[ (1+B)- (3+B)x ]\; {d \over d
x} -
[\; (B+2)^{2} + \omega^{2} \; ]\; F =0\; , \label{12.11c}
\end{eqnarray}

\noindent with
\begin{eqnarray}
\gamma = 1 + B \; , \qquad \alpha + \beta = 2 + B \;, \qquad
\alpha \beta = { (B+2)^{2} + \omega ^{2} \over 4} \; , \nonumber
\end{eqnarray}

\noindent that is
\begin{eqnarray}
B= \pm \;i\; k \; , \qquad \gamma = 1 + B\; , \qquad
 \alpha = {B+2 - i \omega \over 2} \;, \qquad \beta = { B+2 + i
\omega \over 2} \; , \label{12.12a}
\end{eqnarray}

\noindent and
\begin{eqnarray}
E (t,r,z) = \sinh^{2} r \;\; \cosh^{B} r \;\; F(\alpha,
\beta, \gamma, \cosh^{2} r )\; e^{-i (\omega t - k z) }\; .
\label{12.12b}
\end{eqnarray}

Evidently, above constructed  solutions (\ref{12.7a}) --
(\ref{12.7c}) can be obtained from the general relations
(\ref{12.12a}) -- (\ref{12.12b}),  if one demands   $\alpha = 0$ or  $\beta = 0$.
However,
no physical ground exists to impose such (polynomial) restrictions in the case of Lobachevsky space.
Instead, the complete electromagnetic basis should include waves  spreading along   $z$,  with  real parameters
 $k$.

\section{ Solutions at  $k=0$ in space $H_{3}$}

Now let us turn to eqs. (\ref{11.5}) with
$k=0$:
\begin{eqnarray}
({d \over dr } + {\cosh r \over \sinh r} +{\sinh r \over \cosh r }) f_{1}
+
 {im \over \sinh r}\; f_{2} = 0 \; ,
\nonumber
\\
f_{1} ={im \over \omega \sinh r}\; f_{3} \; , \nonumber
\\
 f_{2} =-{1 \over \omega} ({d \over dr} + {\sinh r \over
\cosh r} )\; f_{3} \; , \nonumber
\\
-\omega f_{3} + ({d \over dr} + {\cosh r \over \sinh r} )\; f_{2} -
{im \over \sinh r} f_{1} = 0 \; . \label{13.3}
\end{eqnarray}

\noindent The first  and fourth equations gives
\begin{eqnarray}
({d \over dr } + {\cosh r \over \sinh r} +{\sinh r \over \cosh r }) {im
\over \omega \sinh r}\; f_{3} +
 {im \over \sinh r}\; (-{1 \over \omega} ({d \over dr} + {\sinh r \over
\cosh r} )\; f_{3}) = 0 \; , \nonumber
\\
-\omega f_{3} + ({d \over dr} + {\cosh r \over \sinh r} )\; (-{1 \over
\omega} ({d \over dr} + {\sinh r \over \cosh r} )\; f_{3}) - {im \over
\sinh r} {im \over \omega \sinh r}\; f_{3} = 0 \; . \label{13.4}
\end{eqnarray}

\noindent they reduces respectively to the identity
$0
\equiv 0$ and
\begin{eqnarray}
f_{3}(r) = {1 \over \cosh r}\; E (r) \; ; \qquad
{d^{2} E \over dr^{2}} + {1 \over \sinh r \cosh r}\; {dE \over dr} + (
\omega^{2} - {m^{2} \over \sinh^{2} r} ) E = 0\; . \label{13.7}
\end{eqnarray}

\noindent
In variable $y = - \sinh^{2} r $ it is rewritten as
\begin{eqnarray}
- 4y(1-y) {d ^{2} \over dy^{2}} \; E -4 ( 1 -y) {d \over d y} \; E
+ (\omega^{2} + { m^{2} \over y}) \; E = 0 \;,
\nonumber
\label{13.9}
\end{eqnarray}

\noindent that is solved in hypergeometric functions
$E = y^{a} (1-y)^{b} Y (y)$:
\begin{eqnarray}
4y (1-y) {d^{2} \over dy^{2}} Y + 4[ 1 +2a -(2a+2b+1)y]\; {d \over
d y}Y + \nonumber
\\
+ \left [ -4(a+b)^2-\omega^{2} +(4a^{2}-m^{2}){1 \over y} +
4b(b-1) {1 \over 1-y} \right ] Y=0 \label{13.11}
\end{eqnarray}

\noindent Requiring
\begin{eqnarray}
m= \pm 2a\; , \qquad b = 1 \; , \qquad b = 0 \; , \label{13.13}
\end{eqnarray}

\noindent we get an equation of hypergeometric type with
\begin{eqnarray}
\gamma = 1 + 2a \; , \qquad \alpha + \beta = 2a + 2b \;, \qquad
\alpha \beta = { 4(a+b)^{2} + \omega ^{2} \over 4} \; , \nonumber
\end{eqnarray}

\noindent that is
\begin{eqnarray}
\gamma = 1 + 2a \; , \qquad \alpha = a+b \mp{ i \omega \over 2} \;, \qquad \beta = a+b \pm{ i
\omega \over 2} \; . \label{13.14}
\end{eqnarray}

\section{ Solutions with arbitrary $m,k$ in space  $H_{3}$}

Now let us consider radial equations in general case (\ref{11.5});
the first equation in (\ref{11.5}) turns to be the identity $0=0$ when three remaining hold:
\begin{eqnarray}
-\omega f_{1} = {ik \over \cosh r}\; f_{2} - {im \over \sinh r}\;
f_{3} \; , \nonumber
\\
-\omega f_{2} = ({d \over dr} + {\sinh r \over \cosh r} )\; f_{3} -
{ik \over \cosh r} f_{1} \; , \nonumber
\\
-\omega f_{3} = - ({d \over dr} + {\cosh r \over \sinh r} )\; f_{2} +
{im \over \sinh r} f_{1} \; ; \nonumber
\end{eqnarray}

\noindent with substitutions
\begin{eqnarray}
f_{2} = {1 \over \sinh r} \; F_{2}\; , \qquad f_{3} = {1 \over \cosh r} \; F_{3}\; ,
\nonumber
\label{11.13}
\end{eqnarray}

\noindent one obtains
\begin{eqnarray}
-\omega \; f_{1} = i \; { k \; F_{2} -
 m \; F_{3} \over \sinh r \cosh r } \; ,
\nonumber
\\
-\omega \; {F _{2} \over \sinh r} = {1 \over \cosh r} \; {d F_{3}
\over dr} - {ik \over \cosh r} \; f_{1} \; , \nonumber
\\
-\omega \; {F_{3} \over \cosh r } = - {1 \over \sinh r} \; {d F_{2}
\over dr} + {im \over \sinh r}\; f_{1} \; . \label{11.14}
\end{eqnarray}

\noindent After excluding of $f_{1}$ we get
\begin{eqnarray}
  ( {\omega \over \cosh r} \; {d \over dr}
     + {k m \over \sinh r \cosh^{2} r }
   ) \; F_{3}
    + ( {\omega^{2} \over \sinh r} - {k^{2} \over \sinh r \cosh^{2} r } )\; F _{2} = 0 \; ,
\nonumber
\\
 ( {\omega \over \sinh r} \; {d \over dr} -
{km \over \cosh r \sinh^{2} r} ) \; F_{2} - ( { \omega ^{2} \over \cosh r } -{ m ^{2} \over \cosh r \sinh^{2} r}
     ) \; F_{3} = 0 \; .
\label{14.1}
\end{eqnarray}

\noindent In variable
$\cosh\;  2r -1 = 2y$, eqs. (\ref{14.1}) take the form
\begin{eqnarray}
(2 \omega {d \over dy} - {km \over y (1 + y) }) \; F_{2} +
 ( - { \omega^{2} \over1 + y} + {m^{2} \over y (1+y)})\; F_{3}
= 0\; , \nonumber
\\
(2 \omega {d \over dy} + {km \over y (1 + y) })\; F_{3} + ( + {
\omega^{2} \over y} - {k^{2} \over y (1+y)})\; F_{2} = 0 \; .
\label{14.4}
\end{eqnarray}

Let us introduce new functions by means of a linear transformation (with unit determinant
$\alpha N - \beta M = 1$ and numerical parameters)
\begin{eqnarray}
F_{2} = \alpha \; G_{2} + \beta \; G_{3} \; , \nonumber
\\
F_{3} = M \; G_{2} + N \; G_{3} \; ; \label{14.5}
\end{eqnarray}

\noindent
 and inverse one given by
\begin{eqnarray}
G_{2} = N \; F_{2} - \beta \; F_{3} \; , \nonumber
\\
G_{3} = - M \; F_{2} + \alpha \; F_{3}\; . \label{14.6}
\end{eqnarray}

\noindent Combining equations in  (\ref{14.4}),  we get
\begin{eqnarray}
 2 \omega {d \over dy} \; G_{2} - N \; {km \over y (1 + y) }\; F_{2}
  +
 N \; ( - { \omega^{2} \over1 + y} + {m^{2} \over y (1+y)})\; F_{3} -
\nonumber
\\
 - \beta \; {km \over y (1 + y) } \; F_{3} -
\beta \; ( + { \omega^{2} \over y} - {k^{2} \over y (1+y)})\;
F_{2} = 0 \; , \nonumber
\\[3mm]
2 \omega {d \over dy} \; G_{3} + M\; {km \over y (1 + y) } \;
F_{2}
 -
M \; ( - { \omega^{2} \over1 + y} + {m^{2} \over y (1+y)})\; F_{3}
+ \nonumber
\\
+ \alpha \; {km \over y (1 + y) }\; F_{3} + \alpha ( + {
\omega^{2} \over y} - {k^{2} \over y (1+y)})\; F_{2} = 0 \; .
\nonumber
\\
\label{14.8}
\end{eqnarray}

\noindent
Expressing   $F_{2},F_{3}$ through  $G_{2}, G_{3}$ according to (\ref{14.5}):
\begin{eqnarray}
2 \omega {dG_{2} \over dy} + \left [ - ( N \alpha + \beta M) \;
{km \over y(1+y)} + NM \; { - \omega^{2} y + m^{2} \over y(1 + y)}
- \beta \alpha\; { \omega^{2}(1+y) - k^{2} \over y(1+y) } \;
\right ] \; G_{2} + \nonumber
\\[4mm]
+ \left [ - 2 N \beta \; {km \over y(1+y)} + N^{2} {-\omega^{2} y
+ m^{2} \over y(1 + y) } - \beta ^{2} \; {\omega^{2}(1+y) - k^{2}
\over y(1-y)} \; \right ] \; G_{3} = 0 \; , \label{14.11}
\end{eqnarray}

\begin{eqnarray}
2 \omega {dG_{3} \over dy} + \left [
  ( M \beta + \alpha N ) \; {km \over y(1+y)} -
NM \; { -\omega^{2}y + m^{2} \over y(1 + y)} + \beta \alpha\;
{\omega^{2}(1+y) - k^{2} \over y(1+y)} \; \right ] \; G_{3} +
\nonumber
\\[4mm]
+ \left [ 2 M \alpha \; {km \over y(1+y)} - M^{2} \; {- \omega^{2}
y + m^{2} \over y(1 + y) } + \alpha ^{2} \; {\omega^{2}(1+y) -
k^{2} \over y(1+y) } \; \right ] \; G_{2} = 0 \; , \label{14.12}
\end{eqnarray}

Let us detail the coefficients at  $G_{3}$ and $G_{2}$:
\begin{eqnarray}
\left [ - 2 N \beta \; {km \over y(1+y)} + N^{2} {-\omega^{2} y +
m^{2} \over y(1 + y) } - \beta ^{2} \; {\omega^{2}(1+y) - k^{2}
\over y(1-y)} \; \right ] \; G_{3} = \nonumber
\\
= {1 \over y (1+y) } \; [\; -2 km \; N \beta - \omega^{2} (N^{2} +
\beta^{2}) \; y + N^{2} \; m^{2} + \beta^{2} \; k^{2} \; ]\; ,
\nonumber
\\[3mm]
\left [ 2 M \alpha \; {km \over y(1+y)} - M^{2} \; {- \omega^{2} y
+ m^{2} \over y(1 + y) } + \alpha ^{2} \; {\omega^{2}(1+y) - k^{2}
\over y(1+y) } \; \right ] \; G_{2} = \nonumber
\\
= {1 \over y (1+y) }\; \; [\; 2 km \; M \alpha + \omega^{2} (M^{2}
+ \alpha^{2}) \; y - M ^{2} \; m^{2} - \alpha^{2} \; k^{2} \; ]\;
\nonumber
\end{eqnarray}

\noindent Redundant singularities will be excluded if
\begin{eqnarray}
N^{2} + \beta^{2} = 0 \; , \qquad M^{2} + \alpha^{2} = 0 \; ,
\label{14.13}
\end{eqnarray}

\noindent With additional assumption of unitarity for transformations (\ref{14.5})  and (\ref{14.6})
\begin{eqnarray}
  \alpha \; G_{2} + \beta \; G_{3} = \cos A \; G_{2} + i \sin A \; G_{3} \; ,
\nonumber
\\[2mm]
 M \; G_{2} + N \; G_{3} = i \sin A \; G_{2} + \cos A \; G_{3} \; ,
\label{14.14}
\end{eqnarray}

\noindent we get
$$
 N \alpha + \beta M = \cos 2A \; , \qquad 2N\beta = i\; \sin 2A\; , \qquad 2M \alpha = i\; \sin 2A \; ,
 $$
 $$
  \alpha \beta = i\; \sin A \cos A ={i\over 2} \sin 2A \; ,
\qquad
 NM = i\; \sin A \cos A = {i\over 2} \sin 2A \; ,
$$
$$
N^{2} = \cos^{2} A \; , \qquad \beta^{2} = - \sin ^{2} A \; ,
\qquad M^{2} = - \sin^{2} A\;, \qquad \alpha^{2} = \cos^{2} A \; ;
$$

\noindent correspondingly  eqs. (\ref{14.11}) and (\ref{14.12}) give
\begin{eqnarray}
2 \omega {dG_{2} \over dy} + \left [ - \cos 2A {km \over y(1+y)} +
{i \over 2}\; \sin 2A \;
 { - \omega^{2} y + m^{2} -
  \omega^{2}(1+y) + k^{2} \over y(1+y) } \; \right ] \; G_{2} +
\nonumber
\\
+ \left [ - i\; \sin 2A \; {km \over y(1+y)} + \cos^{2} A \; { -
\omega^{2} y + m^{2} \over y(1 + y) } + \sin ^{2} A\;
{\omega^{2}(1+y) - k^{2} \over y(1+y)} \; \right ] \; G_{3} = 0 \;
, \nonumber
\end{eqnarray}

\begin{eqnarray}
2 \omega {dG_{3} \over dy} + \left [
  \cos 2A {km \over y(1+y)} - {i \over 2} \; \sin 2A \; { -\omega^{2}y + m^{2} -
\omega^{2}(1+y) + k^{2} \over y(1+y)} \; \right ] \; G_{3} +
\nonumber
\\
+ \left [ i\; \sin 2A \; {km \over y(1+y)} + \sin^{2} A \; {-
\omega^{2} y + m^{2} \over y(1 + y) } + \cos ^{2} A \;
{\omega^{2}(1+y) - k^{2} \over y(1+y) } \; \right ] \; G_{2} = 0
\; , \nonumber
\end{eqnarray}

With additional requirement
\begin{eqnarray}
A= \pi / 4 \; , \qquad \cos^{2}A = \sin^{2} A = {1 \over 2} \; ,
\qquad \sin 2A = 1\; , \qquad \cos 2A = 0 \; ; \label{14.15}
\end{eqnarray}

\noindent the system becomes  much more simple
\begin{eqnarray}
2 \omega {dG_{2} \over dy} +i
 { - \omega^{2} y + m^{2} -
  \omega^{2}(1+y) + k^{2} \over 2 y(1+y) } \; G_{2} +
{ - 2i km - \omega^{2} y + m^{2} +
    \omega^{2}(1+y) - k^{2} \over 2y(1+y)} \; G_{3} = 0 \; ,
\nonumber
\end{eqnarray}

\begin{eqnarray}
2 \omega {dG_{3} \over dy} - i
  { -\omega^{2}y + m^{2} -
\omega^{2}(1+y) + k^{2} \over 2 y(1+y)} \; G_{3} + { 2ikm
 - \omega^{2} y + m^{2} +
  \omega^{2}(1+y) - k^{2} \over2 y (1+y) } \; G_{2} = 0 \; ,
\nonumber
\end{eqnarray}

\noindent or
\begin{eqnarray}
\left (2 \omega {d \over dy} -
 i{ \omega^{2} y +
  \omega^{2}(1+y) - m^{2} - k^{2} \over 2 y(1+y) } \right ) \; G_{2} +
{ \omega^{2} + (m-i k)^{2} \over 2y(1+y)} \; G_{3} = 0 \; ,
\nonumber
\end{eqnarray}

\begin{eqnarray}
\left (2 \omega {d \over dy} +
  i{\omega^{2}y +
\omega^{2}(1+y) - m^{2} - k^{2} \over 2 y(1+y)} \right ) \; G_{3}
+ { \omega^{2} + (m+ik)^{2} \over 2y (1+y) } \; G_{2} = 0 \; .
\label{14.16}
\end{eqnarray}

From whence it follows that
\begin{eqnarray}
G_{3}=-2 \omega {2 y (1+y)\over \omega^{2}+(m- i k)^{2}}
{dG_{2}\over dy}+i{ \omega^{2}y+\omega^{2}(1+y)-m^{2}-k^{2}\over
\omega^{2}+(m-i k)^{2}}G_{2},\nonumber
\\[4mm]
4 y (1 + y){d^{2}G_{2}\over dy^{2}}+4 (1 + 2 y){dG_{2}\over dy}+
 \left ( -2 i \omega +\omega^{2 }-{ m^{2}\over y(1+y)}-{
m^{2}+k^{2}\over 1+y} \right ) \; G_{2}=0  \; .
\nonumber
\\
\label{14.18}
\end{eqnarray}

After changing the variable  $ y $ to $ -y$ eq.  (\ref{14.18}) reads
\begin{eqnarray}
4 y (1 - y){d^{2}G_{2}\over dy^{2}}+4 (1 - 2 y){dG_{2}\over dy}-
\nonumber
\\
- \left ( -2 i \omega +\omega^{2 }+{ m^{2}\over y(1-y)}-{
m^{2}+k^{2}\over 1-y} \right ) \; G_{2}=0 \label{14.19}
\end{eqnarray}

Making substitution
\begin{eqnarray}
G_{2} = y^{A} (1-y)^{B} G (y), \nonumber
\nonumber
\end{eqnarray}

\noindent we  get
\begin{eqnarray}
4y (1-y) {d^{2} G\over dy^{2}} + 4[ 1 +2A -(2A+2B+1+1)y]\; {d G
\over d y} -\nonumber
\\
- \left [-\omega (2 i-\omega) +4(A+B)(A+B+1)-{4 A^{2}-m^{2} \over
y}- {4 B^{2}+k^{2} \over 1-y} \right ] G=0 \; . \label{14.21}
\end{eqnarray}

\noindent With additional requirements
\begin{eqnarray}
m= \pm 2A\; , \qquad k= \pm 2 i B\; .
\label{14.23}
\end{eqnarray}

\noindent it  becomes an equation of hypergeometric type
\begin{eqnarray}
4y (1-y) {d^{2} G\over dy^{2}} + 4[ 1 +2A -(2A+2B+1+1)y]\; {d G
\over d y} -\nonumber
\\
- \left[-\omega (2 i-\omega) +4(A+B)(A+B+1)\right ] G=0 \; ,
\label{8.23}
\end{eqnarray}
\begin{eqnarray}
\gamma = 1 + 2A \; , \qquad \alpha + \beta = 2A + 2B+1 \;,
\nonumber
\\
\alpha \beta = { -\omega (2 i -\omega) +4(A+B)(A+B+1) \over 4} \;
, \nonumber
\end{eqnarray}

\noindent that is
\begin{eqnarray}
\alpha =A+B-{i\omega\over 2}  \;, \qquad \beta =
 A+B+1+{i\omega\over 2}. \label{8.24}
\end{eqnarray}

In the same manner, from eqs.  (\ref{14.16}) it follows
\begin{eqnarray}
G_{2}=-2 \omega {2 y (1+y)\over \omega^{2}+(m+i k)^{2}}
{dG_{3}\over dy}-i{ \omega^{2}y+\omega^{2}(1+y)-m^{2}-k^{2}\over
\omega^{2}+(m+i k)^{2}}G_{3},\nonumber
\\[4mm]
 2 \omega {d \over dy}
     \;
\left(-2 \omega {2 y (1+y)\over \omega^{2}+(m+i k)^{2}}
{dG_{3}\over dy}-i{ \omega^{2}y+\omega^{2}(1+y)-m^{2}-k^{2}\over
\omega^{2}+(m+i k)^{2}}G_{3} \right)- \nonumber
\\
-i{ \omega^{2}y + \omega^{2}(1+y) - m^{2} - k^{2} \over 2
y(1+y)}\; (-2 \omega {2 y (1+y)\over \omega^{2}+(m+i k)^{2}}
{dG_{3}\over dy}-\nonumber
\\
 -i{ \omega^{2}y+\omega^{2}(1+y)-m^{2}-k^{2}\over
\omega^{2}+(m+i k)^{2}}G_{3} )+ { \omega^{2} + (m-i k)^{2} \over
2y (1+y) } \; G_{3} = 0 \; , \label{14.28}
\end{eqnarray}

\noindent or
\begin{eqnarray}
4 y (1 + y){d^{2}G_{3}\over dy^{2}}+4 (1 + 2 y){dG_{3}\over dy}+
\nonumber
\\
+ \left ( 2 i \omega +\omega^{2 }-{ m^{2}\over y(1+y)}-{
m^{2}+k^{2}\over 1+y} \right ) \; G_{3}=0 \label{14.29}
\end{eqnarray}

After changing the variable $ y$ to $-y$
\begin{eqnarray}
4 y (1 - y){d^{2}G_{3}\over dy^{2}}+4 (1 - 2 y){dG_{3}\over dy}-
\nonumber
\\
- \left ( 2 i \omega +\omega^{2 }+{ m^{2}\over y(1-y)}-{
m^{2}+k^{2}\over 1-y} \right ) \; G_{3}=0 \; . \label{14.30}
\end{eqnarray}

\noindent and with the substitution
\begin{eqnarray}
G_{3} = y^{A} (1-y)^{B} G (y) \; ,
\nonumber
\end{eqnarray}

\noindent we get
\begin{eqnarray}
4y (1-y) {d^{2} G\over dy^{2}} + 4[ 1 +2A -(2A+2B+1+1)y]\; {d G
\over d y} -\nonumber
\\
- \left [\omega (2 i+\omega) +4(A+B)(A+B+1)-{4 A^{2}-m^{2} \over
y}- {4 B^{2}+k^{2} \over 1-y} \right ] G=0 \; . \label{14.32}
\end{eqnarray}

\noindent With the help of
additional restriction
\begin{eqnarray}
m= \pm 2A\; , \qquad k= \pm 2 i B\; . \label{14.34}
\end{eqnarray}

\noindent the latter reads as an equation of hypergeometric type
\begin{eqnarray}
4y (1-y) {d^{2} G\over dy^{2}} + 4[ 1 +2A -(2A+2B+1+1)y]\; {d G
\over d y} -\nonumber
\\
- \left[\omega (2 i+\omega) +4(A+B)(A+B+1)\right ] G=0 \; ,
\label{14.35}
\end{eqnarray}
\begin{eqnarray}
c= 1 + 2A \; , \qquad a + b = 2A + 2B+1 \;, \nonumber
\\
ab = { \omega (2 i +\omega) +4(A+B)(A+B+1) \over 4} \; , \nonumber
\end{eqnarray}

\noindent that is
\begin{eqnarray}
a = A+B+1-{i\omega\over 2} \;, \qquad b = A+B+{i\omega\over 2} .
\label{14.36}
\end{eqnarray}

Let us find a relative factor in two functions
$G_{2}$ and $G_{3}$. Starting from
\begin{eqnarray}
G_{2} = M_{2} \; y^{\mid m \mid/2} \; (1- y )^{\mid k \mid /2 i
}\; F ( -n , \; n +1 + \mid m \mid + {\mid k \mid\over i}, \; \mid
m \mid + 1 ; \; y ) = \nonumber
\\[3mm]
= M_{2} \; y^{(c-1)/2} (1-y)^{(a+b-c)/2} F( a , b , c , y) \; .
\label{14.38}
\end{eqnarray}

\noindent
Then
\begin{eqnarray}
G_{3} = M_{3} \; y^{\mid m \mid/2 i} \; (1- y )^{\mid k \mid /2 i
}\; F ( -n +1 , \; n + \mid m \mid + {\mid k \mid\over i}, \; \mid
m \mid + 1 ; \; y )= \nonumber
\\[3mm]
= M_{3} \; y^{(c-1)/2} (1-y)^{(a+b-c)/2} F( a +1 , b -1 , c , y)
\; . \label{14.39}
\end{eqnarray}

\noindent and allowing for
\begin{eqnarray}
G_{3} \; [ (m- i k)^{2}+\omega^{2} ] = -4 \omega \; y (1-y) \; {d
G_{2} \over dy} + i [ -m^{2}- k^{2} + \omega^{2} (1 -2y) ] \;G_{2}
\; \; . \label{14.40}
\end{eqnarray}

\noindent we arrive at
\begin{eqnarray}
( m- i k - i\omega ) \; ( m-i k + i\omega) \; {M_{3} \over M_{2} }
 \; F_{3} (y) =
\nonumber
\\
=
 -4 \omega \; [ \;
{ \mid m \mid \over 2} \; (1-y) \; F_{2} (y) -
 {\mid k \mid \over 2 i}\; y\; \; F_{2} (y)
 +
 \nonumber
 \\
 +
 y (1-y) \; {d \over dy} \; F_{2} (y) \; ] +i [ -m^{2}- k^{2} + \omega^{2} (1 -2y) ] \; F_{2} (y) \; .
\label{14.41}
\end{eqnarray}

To find the relative factor it sufficient  to consider the latter  at the point $y=0$  which results in
\begin{eqnarray}
( m- i k - i\omega ) \; ( m-i k + i\omega) \; {M_{3} \over M_{2} }
  =
 -2 \omega \; \mid m \mid -i m^{2}-i k^{2} +i\omega^{2} \; .
\nonumber
\end{eqnarray}

\noindent or
\begin{eqnarray}
i(- i\omega + m-i k ) \; ( i\omega +m-i k ) \; \; {M_{3} \over
M_{2} } = (-i \omega + \mid m \mid -i k ) \; (i\omega + \mid m
\mid + i k ) \; , \nonumber
\end{eqnarray}

\noindent that is
\begin{eqnarray}
M_{2} = i M\; ( -i \omega + m - i k) (i \omega+ m -i k) \; ,
\nonumber
\\
M_{3} =  M\; (-i \omega + \mid m \mid - i k) ( i\omega + \mid m
\mid + i k) \; . \label{14.42}
\end{eqnarray}

These relations  become more simple when separating regions for $m$:
\begin{eqnarray}
m > 0\;, \qquad M_{2} =i M (i \omega - i k + m ) \; , \qquad M_{3}
=  M (i \omega +i k +m ) \; ; \nonumber
\\[3mm]
m < 0\;, \qquad M_{2} =i M (i\omega -i k + m ) \; , \qquad M_{3} =
 M (i\omega +i k  + m ) \; ; \nonumber
\\[3mm]
m=0\; , \qquad M_{2} =  M (k-\omega   )\; , \qquad M_{3} = i
M(k+\omega  ) \; .
\nonumber
\end{eqnarray}

\section{Discussion: on relation to other  formalisms \\ in Maxwell theory}

There exist close relation between the above used covariant technique  in the  complex form of
Riemann-Silberstein-Majorana-Oppenheimer and spinor form of Maxwell theory in general relativity \cite{18}
mainly used in the form of Newman-Penrose formalism of isotropic tetrad \cite{25}:
\begin{eqnarray}
 \left. \begin{array}{ll}
& \qquad i \sigma ^{\alpha }(x)\; [\; \partial /\partial x^{\alpha } \; +
\; \Sigma _{\alpha }(x) \otimes  I  + I \otimes  \Sigma _{\alpha
}(x)\; ]\;
 \xi (x)  = -\; j (x) \; ,  \\[2mm]
 & \qquad  i \bar{\sigma}^{\alpha }(x)\; [ \; \partial /\partial x^{\alpha
}\; +\;
 \bar{\Sigma}_{\alpha }(x) \otimes  I +
I \otimes  \bar{\Sigma}_{\alpha }(x) \; ] \; \eta (x) =- \;
\bar{j}(x) \; ,
\end{array} \right.
\nonumber
\\
\left. \begin{array}{lll}
    \qquad  & \qquad
\nabla^{\beta} F_{\alpha \beta} (x) = -\;j_{\alpha} (x) \; , &\qquad
\epsilon^{\alpha \beta \rho \sigma } \nabla_{\beta} F_{\rho \sigma} (x)  = 0 \; ;
\end{array} \right.
\label{D.1}
\end{eqnarray}
\begin{eqnarray}
 \left. \begin{array}{ll}
 & \qquad  i \bar{\sigma}^{\alpha }(x)\; [\; \partial / \partial x^{\alpha }
\; + \;
 \bar{\Sigma}_{\alpha }(x) \otimes  I + I \otimes  \Sigma _{\alpha }(x)\; ] \;  H(x)  =  \;  \xi \; , \\[2mm]
 & \qquad  i \sigma ^{\alpha }(x)\; [\; \partial /\partial x^{\alpha }\; + \;
\Sigma _{\alpha }(x) \otimes  I +
 I \otimes  \bar{\Sigma}_{\alpha }(x) \;]\; \Delta (x)  =   \; \eta \; ,
\end{array} \right.
\nonumber
\\
\nabla^{\alpha} A_{\alpha} = 0 \; , \qquad \nabla_{\alpha} A_{\beta} - \nabla_{\beta} A_{\alpha} =  \; F_{\alpha \beta} \; ;
\label{D.2}
\end{eqnarray}

\noindent
where   spinor  connections by Infeld -- van der Vaerden are used
\begin{eqnarray}
\sigma ^{\alpha }(x) = \sigma ^{a} \; e^{\alpha }_{(a)}(x) \; ,
\qquad \bar{\sigma }^{\alpha } (x) = \bar{\sigma }^{a} \; e
^{\alpha } _{(a)} (x)\; ,
\nonumber
\\
\Sigma _{\alpha }(x) = {1 \over 2} \; \Sigma ^{ab} \; e^{\beta
}_{(a)} \; \nabla _{\alpha } (e_{(b)\beta }) \;  ,  \qquad
\bar{\Sigma } _{\alpha }(x) = {1 \over 2} \; \bar{\Sigma }^{ab} \;
e^{\beta }_{(x)} \; \nabla _{\alpha } (e_{(b)\beta }) \;   ,
\nonumber
\\
\Sigma ^{a} = {1 \over 4}\; (\;  \bar{\sigma }^{a} \; \sigma ^{b}
\;  - \; \bar{\sigma }^{b} \; \sigma ^{a} \;)\; , \qquad
\bar{\Sigma }^{a}={1  \over  4}  \; ( \; \sigma ^{a}\; \bar{\sigma
}^{b}\; - \;\sigma ^{b} \; \bar{\sigma }^{a}  \; ) \; ;
\end{eqnarray}

\noindent
and electromagnetic bi-spinor  and electric source are  given by
\begin{eqnarray}
  \left | \begin{array}{cc}
\xi  & \Delta  \\
H & \eta  \end{array} \right | = \left | \begin{array}{cc}
 + \; \Sigma^{mn} \;  \sigma^{2} \; F_{mn}    &
+i\; \bar{\sigma}^{l} \sigma^{2} \; A _{l}    \\[2mm]
-i \; \sigma^{l} \sigma^{2}  \;  A _{l} \;    & -\;
\bar{\Sigma}^{mn} \; \sigma^{2} \; F_{mn}
\end{array} \right |\; ;
\nonumber
\\
  j(x) = i \;\sigma^{k} \sigma^{2} \; j_{k}(x) \; , \qquad
   \bar{j}(x) = -i
\;\bar{\sigma}^{k} \sigma^{2} \; j_{k}(x) \; .
\end{eqnarray}

Used in the paper  complex 3-vector approach seems to be a good
alternative to the other  possible techniques. Let us summarize
the content of the paper again.

 Complex formalism
of Riemann - Silberstein - Majorana - Oppenheimer in Maxwell
electrodynamics is extended to the case of arbitrary
pseudo-Riemannian space - time in accordance with the tetrad
recipe of Tetrode - Weyl - Fock - Ivanenko. In this approach, the
Maxwell equations are solved exactly on the background of static
cosmological Einstein model, parameterized by special cylindrical
coordinates and realized as a Riemann space of constant positive
curvature. A discrete frequency spectrum for electromagnetic modes
depending on the curvature radius of space and three parameters is
found, and
 corresponding basis electromagnetic solutions have been constructed explicitly.
 In the case of elliptical model a part of the constructed solutions should be
 rejected by continuity considerations.

Similar treatment is given for Maxwell equations in hyperbolic
Lobachevsky model, the complete basis of  electromagnetic
solutions  in corresponding cylindrical coordinates has been
constructed as well, no quantization of frequencies of
electromagnetic modes  arises.

\section{ Supplement: on the use of matrix complex form of the Maxwell equations
construct   electromagnetic   solutions from
 scalar ones in Minkowski space-time}

The above  matrix form of Maxwell theory :
\begin{eqnarray}
 (-i \partial_{0} + \alpha^{j} \partial_{j} ) \Psi =0 \; , \qquad
\Psi = \left | \begin{array}{c} 0 \\\psi^{1} \\\psi^{2} \\
\psi^{3}
\end{array} \right | \; .
 \label{3.1}
\end{eqnarray}

\noindent being applied to the flat Minkowski model, permits us to develop a simple method of finding
solutions of Maxwell equations on the  base of known solutions of
the scalar massless  equation by  Klein -- Fock -- Gordon. Indeed, in
virtue of the  above commutative  relations  we have  an operator
identity
\begin{eqnarray}
 (-i \partial_{0} + \alpha^{1} \partial_{1} + \alpha^{2}
\partial_{2} + \alpha^{3} \partial_{3} )\; (-i \partial_{0} -
\alpha^{1} \partial_{1} - \alpha^{2} \partial_{2} - \alpha^{3}
\partial_{3} )\;=
 ( - \partial^{2}_{0} + \partial^{2} _{1} +  \partial^{2}_{2} +
\partial^{3}_{3} )\; .
\nonumber
\end{eqnarray}

\noindent Therefore,  taking any special   scalar
solution
 one can immediately  construct four  solutions of
the  Maxwell  equation:
\begin{eqnarray}
( i  \partial_{0}
+ \alpha^{1} \partial_{1} + \alpha^{2} \partial_{2} + \alpha^{3}
\partial_{3} )\;\Phi (x) = \qquad \qquad
\nonumber
\\
=
\left | \begin{array}{rrrr}
i\partial_{0}\Phi &  \quad  \partial_{1}\Phi  &  \quad  \partial_{2}\Phi  &  \quad  \partial_{3} \Phi  \\
-\partial_{1}\Phi &  \quad i\partial_{0}\Phi  &  \quad -\partial_{3}\Phi  &  \quad  \partial_{2} \Phi \\
-\partial_{2}\Phi &  \quad  \partial_{3}\Phi  &  \quad i\partial_{0}\Phi  &  \quad -\partial_{1}\Phi  \\
-\partial_{3}\Phi &  \quad -\partial_{2}\Phi  &  \quad  \partial_{1}\Phi  &  \quad i\partial_{0}\Phi
\end{array}  \right | =\{ \Psi^{0} ,  \Psi^{1}, \Psi^{2} ,\Psi^{3} \}
\; .
\label{3.4}
\end{eqnarray}

Thus,  we have four formal solutions of the free Maxwell equations (let $F_{a} (x) = \partial_{a} \Phi (x)$):
\begin{eqnarray}
\{ \Psi^{0} ,  \Psi^{1}, \Psi^{2} ,\Psi^{3} \} =
\left | \begin{array}{rrrr}
i F_{0}  &  \quad  F_{1} &  \quad  F_{2}  &  \quad  F_{3}   \\
-F_{1}   &  \quad iF_{0} &  \quad -F_{3}  &  \quad  F_{2}   \\
-F_{2}   &  \quad  F_{3} &  \quad iF_{0}  &  \quad -F_{1}   \\
-F_{3}   &  \quad -F_{2} &  \quad  F_{1}  &  \quad iF_{0}
\end{array}  \right | \; .
\label{3.5}
\end{eqnarray}

In general, the function  $\Phi (x)$ is  a  complex-valued.
Relationship  defining all  possible solutions has the structure
\begin{eqnarray}
\lambda _{0} \Psi^{0}  +  \lambda _{1} \Psi^{1} + \lambda _{2} \Psi^{2} + \lambda _{3}\Psi^{3}  =
\left |
\begin{array}{c} 0 \\ {\bf E} + i c{\bf B}
\end{array} \right |.
\label{A.2}
\end{eqnarray}

Substituting these expression   into Maxwell equations and performing calculations needed
  we  can derive  four relations (for more details see \cite{23}):
\begin{eqnarray}
[\;  - \; \lambda_{0} \; \partial _{0} + i \; \lambda_{j}  \; \partial_{j} \; ] \; F_{c} = 0 \; .
\label{A9}
\end{eqnarray}

\noindent
These equations should be analyzed instead of Maxwell equations to construct solutions of  them
in any coordinate system.

It seems reasonable to expect further developments in this matrix based approach to Maxwell theory,
 as a possible base to
explore general method to separate the variables for Maxwell equations in different coordinates
 (see \cite{24}).

\section{Acknowledgements}

Author is grateful  to  participants of seminar of Laboratory of
Theoretical Physics,
 National Academy of Sciences of Belarus for discussion.


This  work was  supported  by Fund for Basic Research of Belarus
 F09K-123.

\end{document}